\documentclass{aa}
\usepackage{graphicx}
\usepackage{natbib}
\usepackage{amssymb}
\usepackage{amstext}
\usepackage{latexsym} 
\usepackage[sumlimits]{amsmath}
\newcommand{\Ha}{H$\alpha$}			%H-alpha
\newcommand{\Hb}{H$\beta$}			%H-beta
\newcommand{\NII}{[N{\sc ii}]}			%[NII]
\newcommand{\HII}{H{\sc ii}}			% HII
\newcommand{\Brg}{Br$\gamma$}			%Br-gamma
\newcommand{\asec}{$^{\prime\prime}$}		% arcsec
\newcommand{\Msolar}{\mbox{\,M$_\odot$}}        % solar mass
\newcommand{\Lsolar}{\mbox{\,L$_\odot$}}        % solar luminosity

\begin{document}
   \title{The H$\alpha$ Galaxy Survey \thanks{
Based on observations made with the William Herschel and Jacobus
Kapteyn Telescopes operated 
on the island of La Palma by the Isaac Newton Group in the Spanish 
Observatorio del Roque de los Muchachos of the Instituto de Astrof\'\i sica 
de Canarias.  The United Kingdom Infrared Telescope is operated by the
Joint Astronomy Centre on behalf of the U.K. Particle Physics and
Astronomy Research Council.
   }}
   \subtitle{II. Extinction and \NII\ corrections to H$\alpha$ fluxes}
   \author{P. A. James
          \inst{1},
          N. S. Shane
          \inst{1},
	  J. H. Knapen 
          \inst{2},
          J. Etherton 
          \inst{1} and 
          S. M. Percival
          \inst{1}
          } \offprints{P. A. James} 
          \institute{Astrophysics Research
	  Institute, Liverpool John Moores University, Twelve Quays
	  House, Egerton Wharf, Birkenhead CH41 1LD, UK \\
	  \email{paj,nss,je,smp@astro.livjm.ac.uk} 
          \and Department of Physics, Astronomy \& Mathematics,
          University of Hertfordshire, Hatfield, 
          Hertfordshire, AL10 9AB, UK\\
          \email{knapen@star.herts.ac.uk}
          }
          \date{Received ; accepted }
   \abstract{

We study the two main corrections generally applied to narrow-band \Ha\ fluxes from
galaxies in order to convert them to star formation rates, namely for
\NII\ contamination and for extinction internal to the galaxy.  From an
imaging study using carefully chosen narrow-band filters, we find the
\NII\ and \Ha\ emission to be differently distributed. Nuclear
measurements are likely to overestimate the contribution of \NII\ to
total narrow-band fluxes. We find that in most star formation regions
in galaxy disks the \NII\ fraction is small or negligible, whereas
some galaxies display a diffuse central component which can be
dominated by \NII\ emission. We compare these results with related
studies in the literature, and consider astrophysical explanations for
variations in the \NII /\Ha\ ratio, including metallicity variations
and different excitation mechanisms.

We proceed to estimate the extinction towards star formation regions
in spiral galaxies, firstly using Br$\gamma$/\Ha\ line ratios.  We
find that extinction values are larger in galaxy nuclei than in disks,
that disk extinction values are similar to those derived from optical
emission-line studies in the literature, and that there is no evidence
for heavily dust-embedded regions emerging in the near-IR, which would
be invisible at \Ha.  The numbers of galaxies and individual regions
detected in Br$\gamma$ are small, however, and we thus exploit optical
emission line data from the literature to derive global \Ha\
extinction values as a function of galaxy type and inclination. In
this part of our study we find only a moderate dependence on
inclination, consistent with broad-band photometric studies, and a
large scatter from galaxy to galaxy.  Typical extinctions are smaller
for late-type dwarfs than for spiral types.  Finally, we show that the
application of the type-dependent extinction corrections derived here
significantly improves the agreement between star formation rates
calculated using \Ha\ fluxes and those from far-infrared fluxes as measured
by {\it IRAS}. This again supports the idea that heavily dust-embedded
star formation, which would be underestimated using the
\Ha\ technique, is not a dominant contributor to the total star
formation rate of most galaxies in the local Universe.

   }
\authorrunning{James et al.}
\titlerunning{H$\alpha$ Galaxy Survey. II.}
\maketitle
%
%________________________________________________________________
\section{Introduction}
\label{sec:intro}

The \Ha\ luminosity observed in spiral and irregular galaxies is
believed to be a direct tracer of the ionisation of the inter-stellar
medium (ISM) by the ultraviolet (UV) radiation which is produced by
young high-mass OB stars.  Since only high-mass ($>$10~\Msolar) and,
therefore, short-lived ($<$20~Myr) stars contribute significantly to
the integrated ionising flux, the \Ha\ emission line thus provides a
nearly instantaneous measure of the star formation rate (SFR),
independent of the previous star formation history.  The two main
advantages of using \Ha\ to detect star formation are the direct
relationship between the nebular line emission and the massive SFR,
and also the high sensitivity.  A small telescope can map star
formation down to low levels at high angular resolution, even in
faint, low surface-brightness galaxies.

The \Ha\ Galaxy Survey (\Ha GS) is a study of the star formation
properties of a representative sample of galaxies in the local
Universe using this technique.  We have imaged 334 nearby galaxies in
both the
\Ha\ line and the $R$-band continuum.  The sample consists of all
Hubble types from S0/a to Im with recession velocities between 0 and
3000~km~s$^{-1}$.  All galaxies were observed with the, now
unfortunately decommissioned, 1.0 metre Jacobus Kapteyn Telescope
(JKT), part of the Isaac Newton Group of Telescopes (ING) situated on
La Palma in the Canary Islands.  The selection and the observation of
the sample are discussed in
\citet{paper1}, hereafter Paper I.  A key element of this project is
to extend the study to fainter galaxies than has been done in previous
large surveys.

The two major limitations of narrow-band \Ha\ imaging are
contamination by the \NII\ line doublet, and uncertainties in the
extinction corrections to be applied to each galaxy.  In this paper we
will investigate refinements to the corrections applied by previous
authors.  The \Ha GS sample contains more faint, dwarf galaxies than
most previous studies, and thus we aim to derive corrections
applicable to both these objects, and the brighter spiral galaxies.

The \NII\ lines are located either side of the 6563\AA\ \Ha\ line, at
6548\AA\ and 6583\AA\, and will be referred to as \NII-6548 and
\NII-6583 in this paper.  The widths of the narrow-band \Ha\ filters
range from 44\AA\ to 55\AA\, thus including both \NII\ lines - as do
almost all narrow-band \Ha\ observations of galaxies.  The \NII-6548
line is weaker than the \NII-6583 line, with
\NII-6583/\NII-6548 $\sim$ 3.  The contribution to the total flux
for both lines together (\NII-total henceforth) causes a scatter when
using \Ha\ measurements to infer SFRs.  In order to improve
calculations of SFRs it is therefore important to be able to correct
for the \NII\ contamination.

Whilst the ratio of \NII\ to \Ha\ in the nuclei of spiral galaxies has
been well studied (e.g., \citealt{keel83,rubin86}) and found to be high
(often greater than 1), \citet{kk83} find that nuclear \NII\ emission
rarely dominates the integrated photometry of galaxies and hence
global corrections for \NII\ contamination are much smaller.  The most
commonly used corrections for entire galaxies
are those derived by \citet{ken83} and \citet{kk83}.
Spectrophotometric
\NII/\Ha\ ratios of individual extragalactic \HII\ regions from the
literature (see \citealt{kk83} and references therein) were compiled from 14
spiral galaxies (mostly of type Sc) and 7 irregular galaxies.  The average
\Ha/(\Ha\ + \NII-total) ratio was found to be fairly constant, spanning the
ranges 0.75 $\pm$ 0.12 for the spirals, and 0.93 $\pm$ 0.05 for the
irregulars (\citealt{ken83}).  In terms of the ratio \NII-total/\Ha\, this
corresponds to a median value of 0.33 for spirals and 0.08 for
irregulars. These values were calculated by finding the
\NII-total/\Ha\ ratio of the brightest \HII\ regions, averaging for
each galaxy and then determining the mean value for spiral and irregular
types.  This implicitly assumes that all
\HII\ regions have the same proportion of \NII-total to \Ha\ emission as
those regions measured.  However, the integrated galaxy
spectroscopy presented by \citet{ken92} shows generally stronger \NII\
emission than the earlier studies.  Mean values of \NII-6583/\Ha\ are
0.75 (st. dev 0.55) for types Sa--Sbc, 0.36(0.16) for types Sc--Sm, and
0.15(0.11) for Im galaxies (omitting the, probably misclassified, galaxy
IC~883 from the Im sample).  Including the \NII-6548 emission would
make the discrepancy larger; however, these ratios were taken from
deblended spectra as the \Ha\ and \NII\ lines were only marginally
resolved.

Similarly, \citet{mcq95} find high mean \NII-total/\Ha\ ratios, of
0.606$\pm$0.432 for 3 normal spirals, 0.909$\pm$1.34 for 12 barred
spirals and 0.838$\pm$0.431 for 5 Type 2 Seyfert galaxies.  The 3 blue
compact dwarf galaxies in their sample have a much lower mean
\NII-total/\Ha\ ratio, 0.104$\pm$0.018, and 3 blue compact galaxies
have an intermediate mean ratio, 0.276$\pm$0.036.  The one LINER in
their sample has very strong \NII\ emission, with \NII-total/\Ha\ $=$ 3.627.  These ratios are
derived from circular apertures of 13.5~arcsec diameter, thus sampling
only the central regions of most of these galaxies.

\citet{jan00} use integrated spectra of 196 nearby galaxies to show
that the principal determinant of \NII/\Ha\ ratio is galaxy luminosity
rather than type. In their study, the \NII/\Ha\ ratios for galaxies
brighter than $M_B = $ -19.5 are in agreement with the values found by
\citet{ken92}, but fainter than this a striking trend is seen
towards much lower values of this ratio.  These fainter galaxies are
dominated in the \citet{jan00} sample by late-type Sd--Im galaxies,
but interestingly the same trend is clearly seen for earlier type
spirals.  At $M_B = $ -16, \NII/\Ha\ ratios of $\sim$0.1 are typical.
A similar trend in log(\NII-total/\Ha\ ) vs galaxy luminosity is found
by \citet{gav04}.

In this paper, a new method of investigating the relative strengths of
\Ha\ and \NII\ is introduced and explored.  This method will enable 
both the integrated ratio over the whole galaxy and the ratios of
individual regions to be calculated.  Particular attention will be
paid to radial variations in the \NII-total/\Ha\ ratio, both for the
practical reason that many literature determinations of this ratio
have been based on measurements of the central regions of galaxies,
and because there are theoretical reasons to expect significant
variation with radius, due to metallicity changes and different
excitation processes.

The major source of systematic error in the conversion of \Ha\ fluxes
to SFRs is due to the effects of extinction within the galaxy being
observed.  There is a great deal of uncertainty as to the exact
magnitude of this extinction and how it varies with galaxy luminosity
and type.  Studies using radio data (which are not affected by
extinction) of large samples of individual \HII\ regions in nearby
galaxies yield mean extinction values ranging from {\it A}(\Ha) = 0.5 mag to
{\it A}(\Ha) = 1.8 mag (e.g. \citealt{caplan86,kaufman87,hulst,caplan96,niklas}).
\citet{ken98} adopts a single correction of 1.1 mag for the effective
\Ha\ extinction of entire galaxies for all Hubble types.  These
samples, however, mainly contain bright galaxies, whereas the \Ha GS
also includes many faint and irregular galaxies. For very small
irregular galaxies, extinction may be systematically lower due to the
smaller quantity of ISM which the \Ha\ flux must traverse.
Alternatively, most of the extinction could be associated with the
star forming region itself, in which case galaxy type and size will be
of lesser importance.  In this paper we will determine the extinction
towards known star formation regions, and search for highly embedded
star formation which may not be apparent through optical observations,
using 3 techniques. The first will present new \Brg\ imaging data for
galaxies in our sample, and use the \Brg /\Ha\ ratio to estimate
extinction towards individual \HII\ regions in these galaxies.
The number of galaxies detected in the \Brg\ line is small, however, 
so a further investigation is presented making use of literature \Ha\
and
\Hb\ line fluxes for a large sample of spiral galaxies.  This sample
enables type-dependent extinction corrections to be derived and
the inclination dependence of extinction to be investigated.  Finally,
a statistical analysis is made of star formation rates for galaxies in the
present sample, using both optical \Ha\ fluxes and far-IR fluxes from
the IRAS survey.  This provides a test of the type-dependent
extinction corrections derived here, and enables a further search for
heavily dust-embedded star formation which may be missed by optical or
even near-IR observations.

The organisation of the rest of this paper is as follows. In Section
\ref{sec:nii}
of this paper we will investigate the fraction of \NII\ emission in
the total \Ha\ + \NII\ flux, how this varies between bulges and disks
of spiral galaxies, and suggest possible reasons for this variation.
In Section \ref{sec:brg}, we use \Brg\ flux measurements from a sample of galaxies
to investigate internal extinction.  In Section \ref{sec:ucm}, we use data
published by the Universidad Complutense de Madrid (UCM) survey
collaboration to investigate the dependences of extinction on galaxy
inclination and morphology. In Section \ref{sec:sfrcomp}, the SFRs calculated using
both \Ha\ and far-infrared (FIR) data are compared, used to check
the type-dependent extinction corrections, and to constrain
deeply-embedded star formation.  Section \ref{sec:conc} summarizes the main
conclusions of this paper.

%__________________________________________________________________

\section{Photometric separation of \NII\ and \Ha}
\label{sec:nii}
%------------------------------------------------------------------------------------
\subsection{Methods}

The \NII\ 6584\AA\ filter (n6584) at the JKT has a narrow
passband (21\AA), which is centred on the \Ha\ line for galaxies with
recession velocities close to 960~km~s$^{-1}$.  If such galaxies are
observed through the n6584 filter, the narrow passband 
virtually excludes ($<10$\% transmission) the \NII\ lines.
Observations taken through the wider 6594\AA\ \Ha\ filter (h6594) in
the JKT filter set include both the
\NII-6583 and the
\Ha\ emission.  Using the transmission values from the scanned
filter profiles on the ING website (see Table \ref{tbl:niigals} for
specific values adopted for each observed galaxy), the
continuum-subtracted images can be scaled and subtracted so as to
produce pairs of images in the \Ha\ and \NII-6583 lines respectively
(the \NII-6548 line is included in the analysis by multiplying
\NII-6583 fluxes by a factor of 4/3). The key point to note in Table
\ref{tbl:niigals} is that the transmission values of the \Ha\ line are very
similar for the two filters, while those for the \NII-6583 line are an
order of magnitude lower in the narrower filter; this difference
enables the separation process to work. The relative strengths of the
\Ha\ and \NII-6583 fluxes can be derived from the separated images 
using aperture photometry.

By applying this method to a range of galaxies of different types, it
is possible to look at variations in the ratio of \Ha\ to \NII\ both
as a function of galaxy type, and of spatial location within each
galaxy.  In the latter case the effects resulting from radial
metallicity gradients, for example, can be investigated.
%------------------------------------------------------------------------------------
\subsection{Observations}

Five hours of service time were granted on the JKT to obtain images
through the n6584 filter for the study of \NII\ emission strengths. The
observing took place in December 2000 and January 2001.  Three
galaxies were observed through the n6584 (3 $\times$ 1200~s
integrations) and $R$-band continuum filters.  \Ha\ data were taken as
part of the main survey.  Two further galaxies were observed through
the n6584 filter during the main \Ha GS observing time.

The galaxies range from spiral types S0/a to Scd and are listed in
Table \ref{tbl:niigals}, along with their recession velocities.
Photometric calibration was obtained by using observations of Landolt
standard stars to define zero-points and airmass corrections for each
night as described in Paper I.

\begin{table*}
\begin{center} 
\begin{small} 
\begin{tabular}{llccccc}
\hline
\hline
Galaxy  & Hubble Type & $v$(km~s$^{-1}$) & $T_{\rm H\alpha,h6594}$ &
$T_{\rm H\alpha,n6584}$ & $T_{\rm [NII],h6594}$ & $T_{\rm [NII],n6584}$ \cr
\hline
UGC~2141   & S0/a        & 987  & 0.4149 & 0.4452 & 0.4626 & 0.0680 \cr
UGC~2210   & SBc         & 1211 & 0.4683 & 0.4298 & 0.3958 & 0.0203 \cr
UGC~855    & SABc        & 1202 & 0.4667 & 0.4320 & 0.3992 & 0.0213 \cr
UGC~8403   & SBcd        & 965  & 0.4084 & 0.4425 & 0.4671 & 0.0756 \cr
UGC~11872  & SABb        & 1150 & 0.4568 & 0.4446 & 0.4171 & 0.0297 \cr
\hline
\end{tabular}
\caption[]{Galaxies observed in \NII\ emission, and the \Ha\ and \NII\ 
transmission values for both lines through the h6594 and n6584 filters.}
\label{tbl:niigals}
\end{small} 
\end{center}
\end{table*}

%------------------------------------------------------------------------------------
\subsection{Producing separate {\rm \Ha\ }and {\rm \NII\ }images}

The $R$-band continuum images were scaled and subtracted from the
n6584 filter images.  The scaling factor was found to be $R$/n6584 =
118.5$\pm$1.8 from photometry of 41 foreground and standard stars.  This is
identical, within the errors, to the value of 118.4$\pm$2.2 obtained by
multiplying spectrophotometric stellar fluxes by the digitised filter
transmission profiles.

The continuum-subtracted images taken through the h6594 filter were
scaled so as to match the photometric zeropoint of the n6584 filter
images.  In the cases where there were large differences in the seeing
conditions in the two images, a Gaussian smoothing was applied to the
image with the better seeing. The continuum-subtracted images were
scaled by the appropriate factors from Table \ref{tbl:niigals} and
subtracted to produce one image mapping the \NII\ emission from the
galaxy and one image containing only the \Ha\ light.

%------------------------------------------------------------------------------------
\subsection{{\rm \NII\ }and {\rm \Ha\ }distributions in individual galaxies}
\label{sec:niifindings}

For each of the 5 galaxies, the ratio of the \NII-6583 to the \Ha\
flux and the distribution of each were investigated for the whole
galaxy by plotting growth curves.  The fluxes were measured in,
typically, 50-80 equally spaced elliptical apertures.  The location of
the galaxy centre was determined by centroiding on the galaxy nucleus
in the $R$-band image, and the ellipticity and position angle of the
apertures were taken from the Uppsala Galaxy Catalogue (UGC,
\citealt{nilson}). The growth curve from the separated \NII-6583 image
was divided by that obtained from the separated \Ha\ image to obtain
the ratio \NII-6583/\Ha\ as a function of distance from the galaxy centre. 

Figures 1 and 2 show this ratio and the two growth curves, normalised
to the maximum \NII-6583/\Ha\ ratio, displayed in the upper frame of
each group.  The middle frame of each group is the uncontaminated \Ha\
image and the bottom image shows the distribution of the \NII-6583
emission.

\paragraph{UGC~2141}

\begin{figure*}
\centering
\includegraphics[angle=0,width=5.9cm]{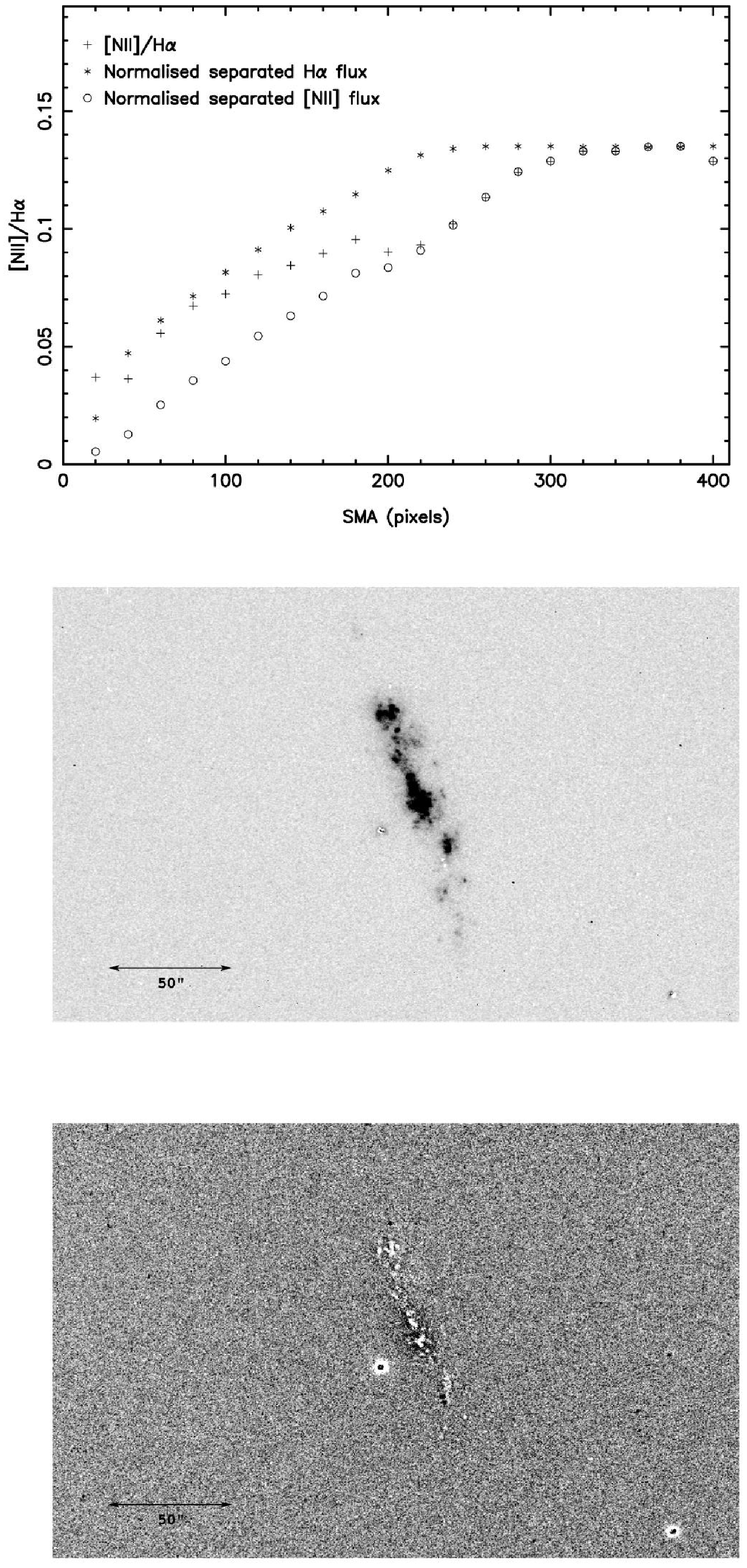}
\includegraphics[angle=0,width=5.9cm]{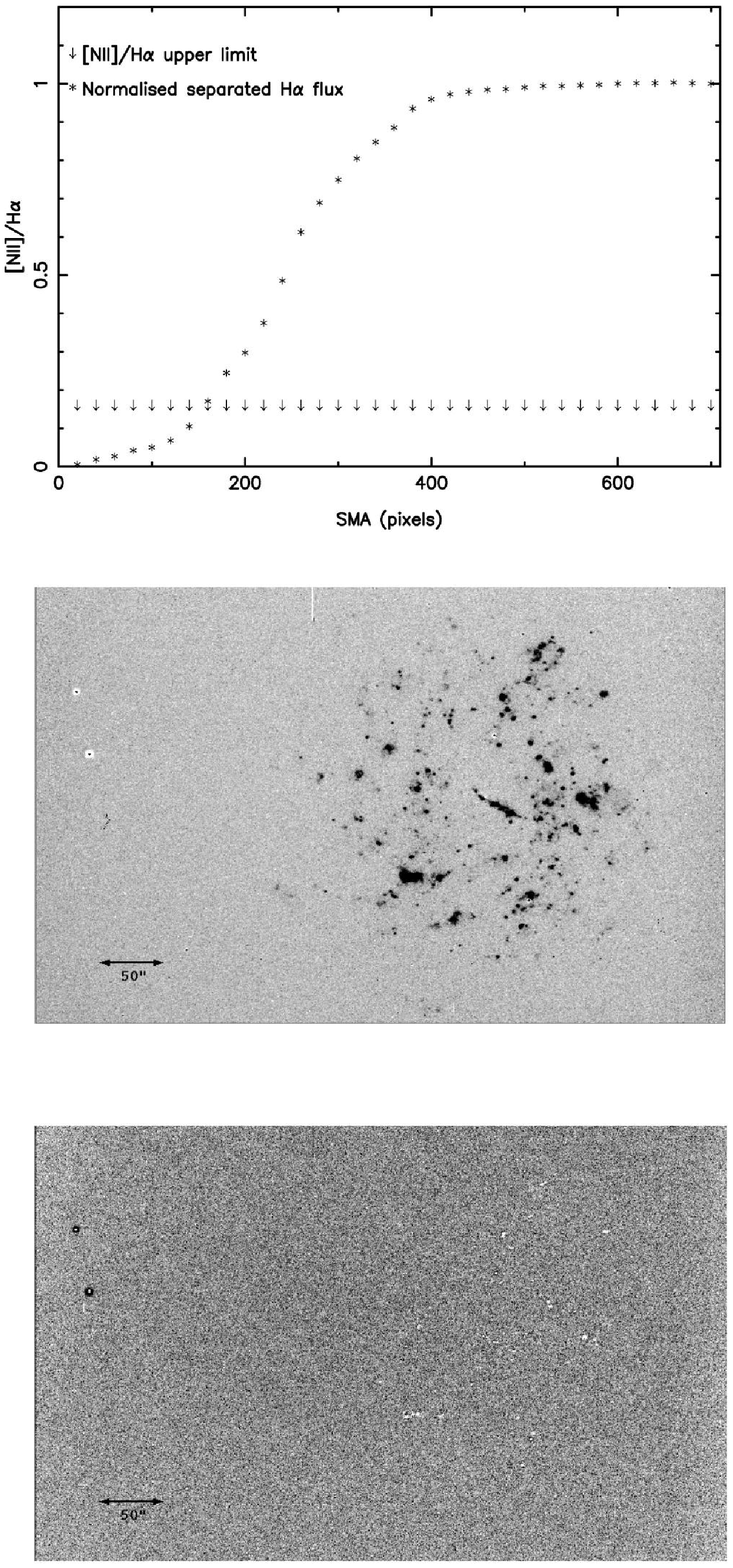}
\includegraphics[angle=0,width=5.9cm]{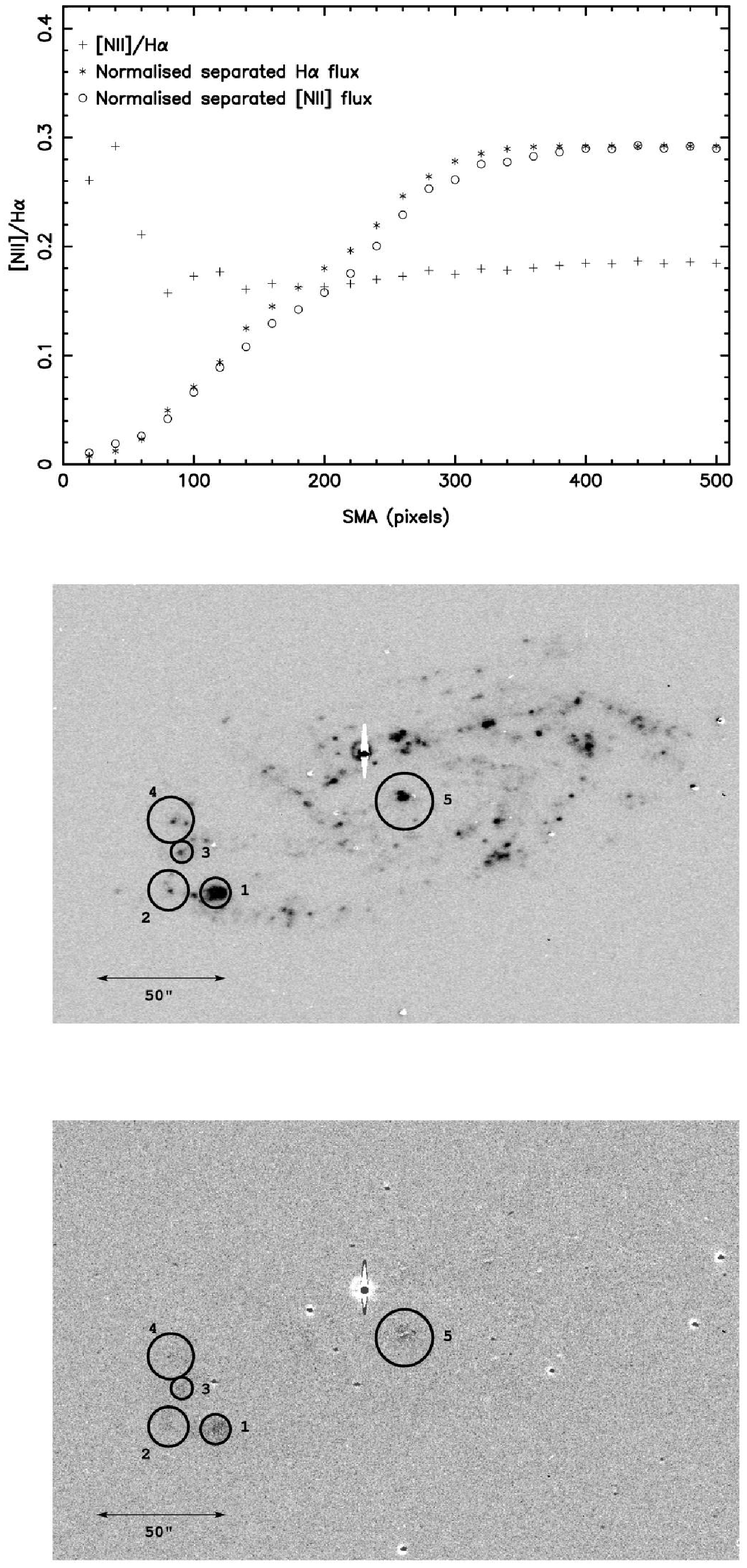}

\caption{UGC~2141 (left), UGC~2210 (centre) and UGC~2855 (right).  In
each case, the top frame shows the growth curves in \Ha ,
\NII-6583 , and in the ratio \NII-6583/\Ha , except for UGC~2210 where
only an upper limit is shown for the \NII-6583/\Ha ratio, as no
\NII\ is detected in this galaxy.  The middle and lower frames show
the separated images in the \Ha\ and \NII\ lines respectively for each
galaxy.  The areas shown in the images are 4.6$\times$2.9~arcmin, or
16$\times$10~kpc for UGC~2141; 9.1$\times$5.7~arcmin, or
37$\times$23~kpc for UGC~2210; and 4.3$\times$2.7~arcmin, or
22$\times$14~kpc for UGC~2855.
}
\label{fig:u2141_u2210_u2855}
\end{figure*}

Figure \ref{fig:u2141_u2210_u2855} (left) shows that in this galaxy, the \NII-6583 and
\Ha\ do not separate into distinct regions as found in some of the
other galaxies.  The \NII-6583 image shows a diffuse mix of positive
and negative readings, showing that the method has not worked
particularly well in this case.  This will be, in part, due to the
large seeing difference between the nights in which the galaxy was
observed through the h6594 filter (2.0~arcsec) and the n6584 filter
(1.0~arcsec).
However, this should not affect the large-scale trends or overall
ratios shown in Fig. \ref{fig:u2141_u2210_u2855}.

The ratio of \NII-6583 to \Ha\ increases in the outer regions of the
galaxy and levels off at a value of 0.13.  Thus, the value of
\Ha/(\Ha\ + \NII-total) is 0.85, which is higher than the mean value suggested
\citet{ken83}, and \NII\ emission is relatively weak in this galaxy.

\paragraph{UGC~2210}
Figure \ref{fig:u2141_u2210_u2855} (centre) shows virtually no \NII\ emission.
The \NII\ growth curve shows no significant departures from zero and
it is only possible to put an upper limit on the
\NII\ to \Ha\ ratio for the whole galaxy.  The upper limits shown in
this figure correspond to the highest value detected in any of the
photometric apertures used.  The 3$\sigma$ upper limit on the
\NII-6583/\Ha\ ratio for the whole galaxy is 0.02, which is equivalent to
a lower limit of 0.97 on \Ha/(\Ha\ + \NII-total).

\paragraph{UGC~2855}
Figure \ref{fig:u2141_u2210_u2855} (right) shows that there is a high
ratio of \NII-6583 to \Ha\ in the nucleus, but this ratio falls significantly in the disk.
There is an arc of \NII-6583 emission, however, located at the eastern end
of one of the spiral arms.

The asymptotic value of \NII-6583/\Ha\ is 0.18.  \Ha/(\Ha\ + \NII-total) is,
therefore, 0.80 for the whole galaxy.

\begin{figure*}
\centering
\includegraphics[angle=0,width=5.9cm]{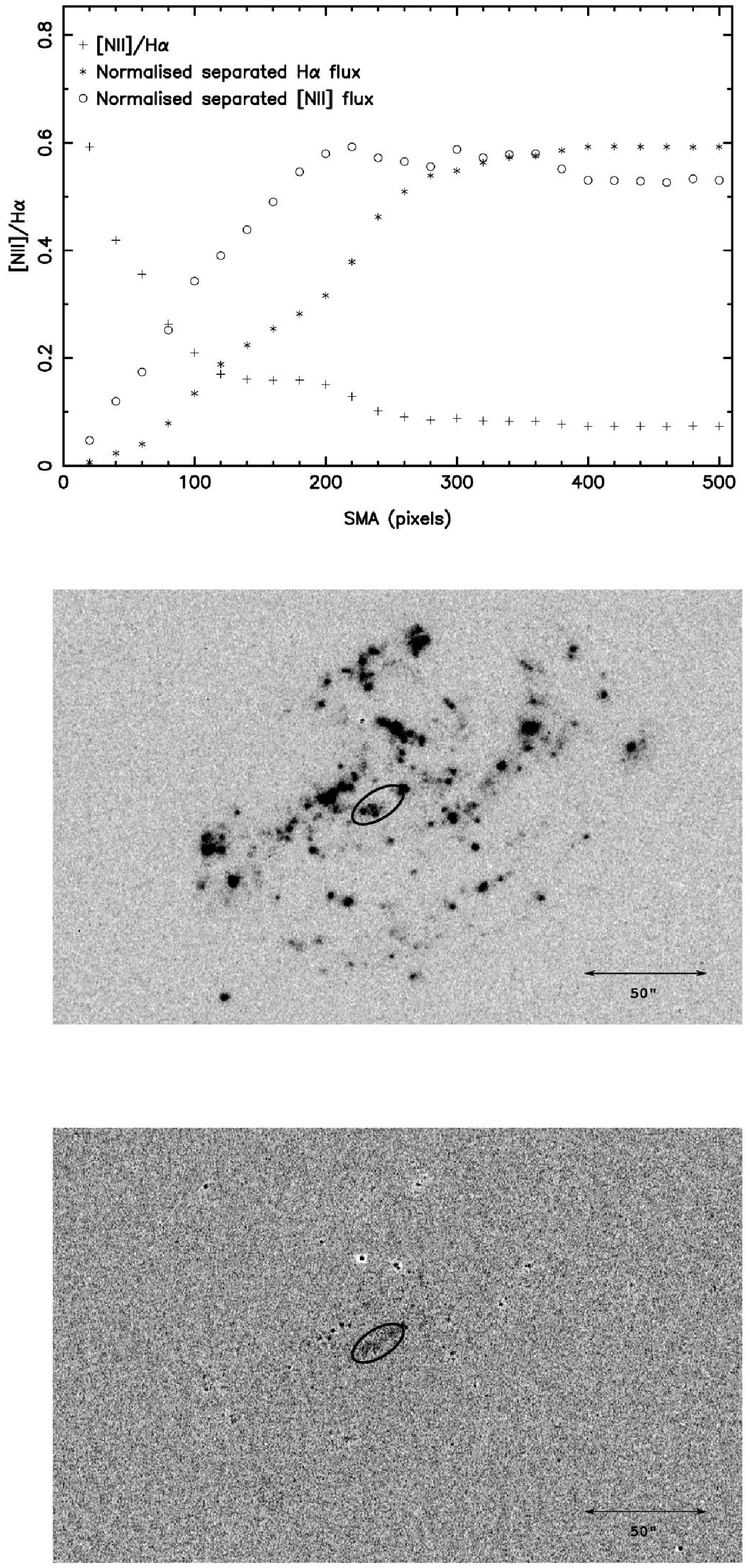}
\includegraphics[angle=0,width=5.9cm]{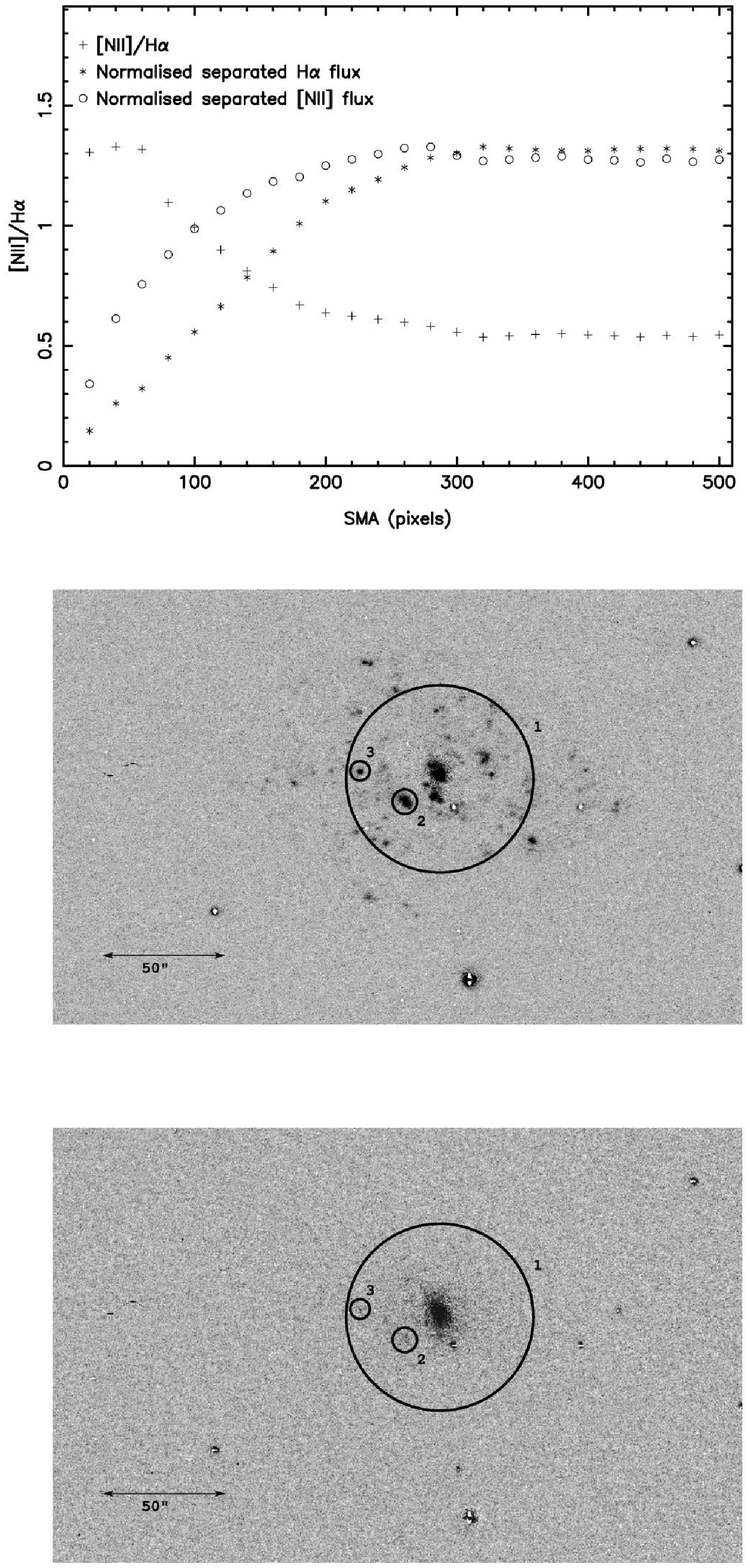}
\caption{UGC~8403 (left) and UGC~11872 (right). The frames are
arranged as for Fig. 1; the areas shown in the images are
4.6$\times$2.9~arcmin, or 26$\times$16~kpc for UGC~8403; and
4.6$\times$2.9~arcmin, or 18$\times$11~kpc for UGC~11872. }
\label{fig:u8403_u11872} 
\end{figure*}

\paragraph{UGC~8403}
Figure \ref{fig:u8403_u11872} (left) shows strong nuclear \NII-6583 emission,
peaking in the centre with an \NII-6583/\Ha\ ratio of 0.59.  This falls
rapidly and there is very little sign of \NII\ in the disk.  The
average ratio for the whole galaxy is just 0.07, which corresponds to
\Ha/(\Ha\ + \NII-total) = 0.91.

\paragraph{UGC~11872}
Figure \ref{fig:u8403_u11872} (right) shows very strong \NII-6583 emission in the
galaxy nucleus, with a ratio of \NII-6583 to
\Ha\ of 1.33 over the central 20$^{\prime\prime}$ in radius.  UGC~11872 
is listed on NED as containing a
low-ionisation nuclear emission region (LINER).  The \NII-6583 emission is
much weaker in the outer regions of the disk, but the total \NII-6583/\Ha\
ratio for the galaxy is still 0.54, easily the largest fraction in the 5
galaxies studied.  The fraction of
\Ha\ to the  \Ha\ + \NII-total flux is 0.58 for the entire galaxy.

The above results are summarised in Table \ref{tbl:niiall}.  The
quoted errors represent the standard deviation of individual points
about the asymptotic values in the \NII/\Ha\ growth curves.

\begin{table*}
\begin{center} 
\begin{small} 
\begin{tabular}{llcc}
\hline
\hline
Galaxy  &  \NII-6583/\Ha       & \Ha/(\Ha\ + \NII-total)  \cr
\hline 
U2141   & 0.133$\pm0.002$      & 0.849$\pm0.002$ \cr
U2210   & $<$0.020             & $>$0.974        \cr
U2855   & 0.185$\pm0.001$      & 0.802$\pm0.001$ \cr
%U3580   & $<$0.034             & $>$0.967        \cr
U8403   & 0.074$\pm0.001$      & 0.910$\pm0.001$ \cr
U11872  & 0.541$\pm0.003$      & 0.582$\pm0.002$ \cr
\hline
\end{tabular}
\caption[]{Ratios of \NII-6583 to \Ha\ and \Ha\ to \Ha\ + \NII-total 
measured for the entire galaxy}
\label{tbl:niiall}
\end{small} 
\end{center}
\end{table*}

For the three galaxies where individual \NII-emitting regions were
visible, the ratio of \NII\ to \Ha\ was investigated separately in
each emission-line region.  Circular or elliptical apertures were
placed around the regions in the positions indicated in Figs.
\ref{fig:u2141_u2210_u2855} and \ref {fig:u8403_u11872}.  The fluxes
were recorded in identical apertures in both the
\Ha-only image and the \NII-6583 image.  The results are combined in
Table \ref{tbl:niiregions}.

\begin{table*}
\begin{center} 
\begin{small} 
\begin{tabular}{llcc}
\hline
\hline
Region &    & \NII-6583/\Ha      & \Ha/(\Ha\ + \NII-total)  \cr
\hline
U2855  & 1  & 0.228$\pm0.010$    & 0.767$\pm0.006$ \cr
       & 2  & 0.520$\pm0.067$    & 0.591$\pm0.028$ \cr
       & 3  & 0.407$\pm0.028$    & 0.649$\pm0.014$ \cr
       & 4  & 0.529$\pm0.073$    & 0.587$\pm0.033$ \cr
       & 5N & 0.289$\pm0.029$    & 0.722$\pm0.017$ \cr
U8403  & N  & 0.536$\pm0.024$    & 0.582$\pm0.010$ \cr
U11872 & 1N & 0.800$\pm0.032$    & 0.484$\pm0.010$ \cr
       & 2  & 0.337$\pm0.013$    & 0.691$\pm0.007$ \cr
       & 3  & 0.277$\pm0.018$    & 0.731$\pm0.011$ \cr
\hline
\end{tabular}
\caption[]{Ratios of \NII-6583 to \Ha\ and \Ha\ to \Ha\ + \NII-total for 
individual regions.  An N next to a region name indicates a 
nuclear region.}
\label{tbl:niiregions}
\end{small} 
\end{center}
\end{table*}

If we were to assume that all \HII\ regions in a galaxy have
approximately the same ratio of \NII\ to \Ha, then the mean value of
\Ha/(\Ha\ + \NII-total) for spiral galaxies, calculated from the individual
regions listed in Table \ref{tbl:niiregions}, would be
0.645$\pm$0.03.  This is in fair agreement with the value quoted in
\citet{ken83} of 0.75$\pm$0.12, which was derived using effectively 
the same assumption.  However, the majority of the above
regions were selected on their \NII\ emission, which can obviously
bias the mean ratio.  The process was
repeated for some of the most luminous regions in the
\Ha-only image.  The \NII\ fluxes for these regions were predicted 
assuming \Ha/(\Ha\ + \NII-total) = 0.75 and should have all been easily
detectable.  With the exception of regions 2 and 3 in
UGC~11872, {\em no} corresponding \NII\ flux could be measured in any of
these areas.  

The 3$\sigma$ upper limit on the \NII\ to \Ha\ ratio
lies between 0.034 and 0.117 for the regions investigated.  Since
these limits are well below the average ratio found from Table
\ref{tbl:niiregions}, the assumption
of equal ratios of \NII\ to \Ha\ in all \HII\ regions can not be
valid.

A further test of this is to remove the regions in which \NII\ was
detected and use a single large aperture to measure the ratio of the
remaining \NII\ to \Ha\ in each galaxy.  The results are presented in
Table \ref{tbl:niiremoved}, and again show that the ratio of \NII\ to
\Ha\ is low outside the most strongly line-emitting regions.

\begin{table*}
\begin{center} 
\begin{small} 
\begin{tabular}{lcc}
\hline
\hline
Galaxy &  \NII-6583/\Ha &  \Ha/(\Ha\ + \NII-total)  \cr
\hline
U2855  & 0.147$\pm0.002$      & 0.836$\pm0.002$        \cr
U8403  & 0.068$\pm0.001$      & 0.917$\pm0.001$        \cr
U11872 & 0.241$\pm0.002$      & 0.757$\pm0.001$        \cr
\hline
\end{tabular}
\caption[]{Ratios of \NII-6583 to \Ha\ and \Ha\ to \Ha\ + \NII-total for 
the remainder of the galaxy after the regions in Table
\ref{tbl:niiregions} have been removed.}
\label{tbl:niiremoved}
\end{small} 
\end{center}
\end{table*}

These data confirm that, whilst the results of \citet{ken83} can be
reproduced for the regions with the strongest \NII\ emission, the
overall strength of \NII\ relative to \Ha\ in the remainder of each of
the 3 galaxies is lower than the value derived by \citet{ken83}.

\subsection{Literature studies of {\rm \NII } emission}

\citet{burbidge} found evidence for the variation in the \NII-total/\Ha\
ratio within spiral galaxies, with the typical ratio being $\sim$1/3
in disk regions, but rising to $\sim$1 (occasionally significantly
higher) in nuclear regions.  They suggested that such variations could
be due to either metallicity gradients, resulting in a higher nitrogen
abundance in the central parts of galaxies, or to a higher electron
temperature in these regions.  \citet{osterbrock} also notes the
ambiguity inherent in interpreting the \NII /\Ha\ ratio, but he notes
that it clearly splits `HII-like' galaxies from those thought to be
powered by active galactic nuclei.  This point is discussed by 
\citet{baldwin}, who demonstrate that higher values of the ratio can
result either from shock heating, or from photoionization with a
power-law spectrum, rather than the near-black-body spectrum expected
from stellar photoionization.  The nearby Sab LINER, M81, provides a
particularly well-studied example of the systematic variations found
in the \NII /\Ha\ ratio.  \citet{stauffer} measured the \NII-6583 /\Ha\ ratio
for 10 disk regions, finding values in the range 0.22--0.45, whereas
the nuclear value of this ratio is found by \citet{filippenko} to be
much larger at 2.3.  Most authors have followed \citet{peimbert} in
ascribing the radial changes in this ratio primarily to the abundance
gradient.

A highly interesting phenomenon of great relevance to the current
study is the occurrence of Extended Nuclear Emission-line Regions
(ENERs) noted in the bulge regions of nearby spiral galaxies by
Devereux and collaborators (\citealt{dev94,dev95,hameed}). These are
areas of diffuse emission, seen through filters transmitting
\Ha $+$ \NII\ light, which do not have the clumpy appearance
associated with star formation regions, and are cospatial with the old
or intermediate-age stellar populations of galaxy bulges.  Such
components have been found in M~31 (\citealt{dev94}), M~81
(\citealt{dev95}), and in 7 out of a sample of 27 Sa--Sab galaxies
studied by \citet{hameed}.  These authors argue strongly that these
extended components, which typically comprise $\sim$30\% of the \Ha
$+$ \NII\ luminosity in these galaxies, are completely unrelated to
star formation activity, and they propose that this emission is either
shock-excited, or photo-ionized by extremely hot post-AGB stars.  The
latter option provides a simple explanation for this emission being
distributed throughout the bulge region.  It is clear that the
central extended  emission found in the present study, most clearly in
UGC~11872, but also from UGC~2855 and UGC~8403, is the same ENER phenomenon
noted by Devereux and collaborators, and the high \NII-6583 /\Ha\
ratio we find throughout these regions lends weight to the suggestion
that these are not excited by star formation.

%-------------------------------------------------------------------
\subsection{{\rm \NII\ }vs {\rm \Ha\ }emission - summary and conclusions}

The method discussed here for photometrically separating the \NII\ and
\Ha\ emission from galaxies can be used both to refine the
\NII\ corrections proposed by \citet{ken83} for future \Ha\ based
studies of star formation activity, and to study the nature of
emission-line sources in galaxies.
The current sample size of this investigation is very small, but even
so, we see a wide variation in the strength and distribution of \NII\
emission in each galaxy. For example, UGC~2210 and UGC~2855 are both Sc
galaxies, and yet their \NII\ profiles are very different. No
correlation is evident here between between \NII\ / \Ha\ ratio and galaxy
luminosity, but this is not surprising given the small sample size and
the scatter in this relation found by \citet{jan00} and \citet{gav04}.
UGC~2141, 2855 and 11872 lie close to the best-fit line for this
correlation as presented by \citet{gav04}, but UGC~8403 has weak
\NII\ emission for a relatively bright spiral galaxy, and UGC~2210 has
weaker \NII\ than any galaxy studied by \citet{gav04}.

The \citet{ken83} correction agrees well with this work for individual
regions selected by their high \NII\ fluxes.  For the remaining
galaxy, and the galaxy as a whole, however, the correction
significantly overestimates the \NII\ contamination effects.

Clearly, further observations through the n6584 filter are required,
particularly of irregular galaxies, before a new set of \NII\
corrections can be derived.  For the SFRs published in Paper I, the
current standard corrections were used for comparability with other
work.  If a value of \Ha/(\Ha\ + \NII) of 0.823, suggested by the mean
of the values for the 5 entire galaxies in Table
\ref{tbl:niiall}, is taken, then the derived SFRs as reported in Paper
I are too low by 7\%.

The photometric separation method is also a good way of investigating
the distribution of \NII\ within each galaxy.  Three out of the 4
galaxies for which positive \NII\ measurements could be made showed
higher \NII/\Ha\ fractions in the metal-rich nuclear regions than in
their younger disks, including diffuse components which we identify
with the ENER phenomenon discussed by Devereux and collaborators
(\citealt{dev94,dev95,hameed}).  The higher nuclear ratios are
consistent with the findings of earlier studies (e.g.
\citealt{burbidge}), and may explain the high values found by
\citet{mcq95}, whose study was based on spectroscopy of central regions
of galaxies.

%__________________________________________________________________

\section{Investigating internal extinction effects using \Brg\ observations}
\label{sec:brg}

\subsection{Methods}

Compared to \Ha, the effective extinction at the wavelength of the
\Brg\ 2.166$\mu$m hydrogen recombination line is lower by a factor of
7.1 (calculated using the extinction curve of \citealt{cardelli}),
making this line an excellent probe of dust-embedded star formation.
Assuming case B recombination (\citealt{osterbrock}), the intrinsic
\Brg/\Ha\ line ratio can be predicted. Comparing the measured and
predicted ratios, we can calculate the excess extinction at \Ha\
compared to \Brg, and hence the total extinction at \Ha, by the
assumption of a standard extinction law.

The \Brg\ line is relatively weak, however, with a flux 104 times
weaker than that of \Ha\ (\citealt{osterbrock}; this assumes
$T=$10,000~K and an electron density $N_{\rm e}=10^4$~cm$^{-3}$). Thus
a 4~metre class telescope is required for \Brg\ observations in
galaxies.  The UFTI (UKIRT Fast Track Imager) camera at UKIRT (United
Kingdom InfraRed Telescope) with the
\Brg\ and \Brg z filters (centred on 2.166$\mu$m and 2.173$\mu$m
respectively) is ideal for this work as the filters can be used for
galaxies with recession velocities up to 4500 km~s$^{-1}$,
encompassing the full range of recession velocities for galaxies in
our sample (0--3000 km~s$^{-1}$). Three nights of observing time
(2001 January 17-19) on UKIRT were awarded for \Brg\ observations of a
subset of the \Ha\ sample.

%------------------------------------------------------------------------------------

\subsection{UKIRT observations and data reduction}
\label{sec:ukirt_obs}
Thirty-nine galaxies were selected from the \Ha\ survey sample,
subject only to the selection criteria of this survey
(Paper I) and the need to sample all spiral and irregular
Hubble types, and excluding the largest galaxies ($D_{\rm 25}>$3.5~arcmin)
because of the field of view of the infrared camera used. Twenty-one of these
were spirals ranging between Hubble types S0/a to Sd inclusive, and 18
were irregulars with a range of absolute magnitudes.  During the
three night run, 22 of the galaxies were observed through one of the two
\Brg\ filters and the $K^{\prime}$ continuum filter.  Given the
difficulties encountered in detecting \Brg\ emission, the decision was
made at the telescope to target preferentially galaxies with the
strongest \Ha\ line emission.

Exposure times varied from 600~s to 3600~s through the \Brg\ filters,
depending on the magnitude of the galaxy.  The \Brg\ filter was used
for galaxies with velocities less than 1000 km~s$^{-1}$, and the \Brg
z filter for those with higher redshifts.  A 300~s exposure was taken
for each galaxy through the $K^{\prime}$ filter (centred on
2.1123$\mu$m) for continuum subtraction.  A dark image was taken for
each observation, and sky flats and standard stars were observed
throughout the night.  Observations of standard stars confirmed that
all three nights were photometric, and were used to calculate the
scaling factors between the two \Brg\ filters and the $K^{\prime}$
filter.  For the 10 stars observed through both the $K^{\prime}$ and
the \Brg\ filters, the mean ratio of detected counts,
$K^{\prime}$/\Brg, was found to be 10.74$\pm$0.10.  A value for
$K^{\prime}$/\Brg z of 7.13$\pm$0.09 was derived from the 8 standard
stars observed through these filters.  Appropriately scaled and
aligned $K^{\prime}$ images were subtracted from each \Brg\ frame to
remove the continuum emission.

%------------------------------------------------------------------------------------

\subsection{\Brg\ detections}
\label{sec:brg_detect}
\begin{figure*}
\centering
\includegraphics[angle=0,width=8.5cm]{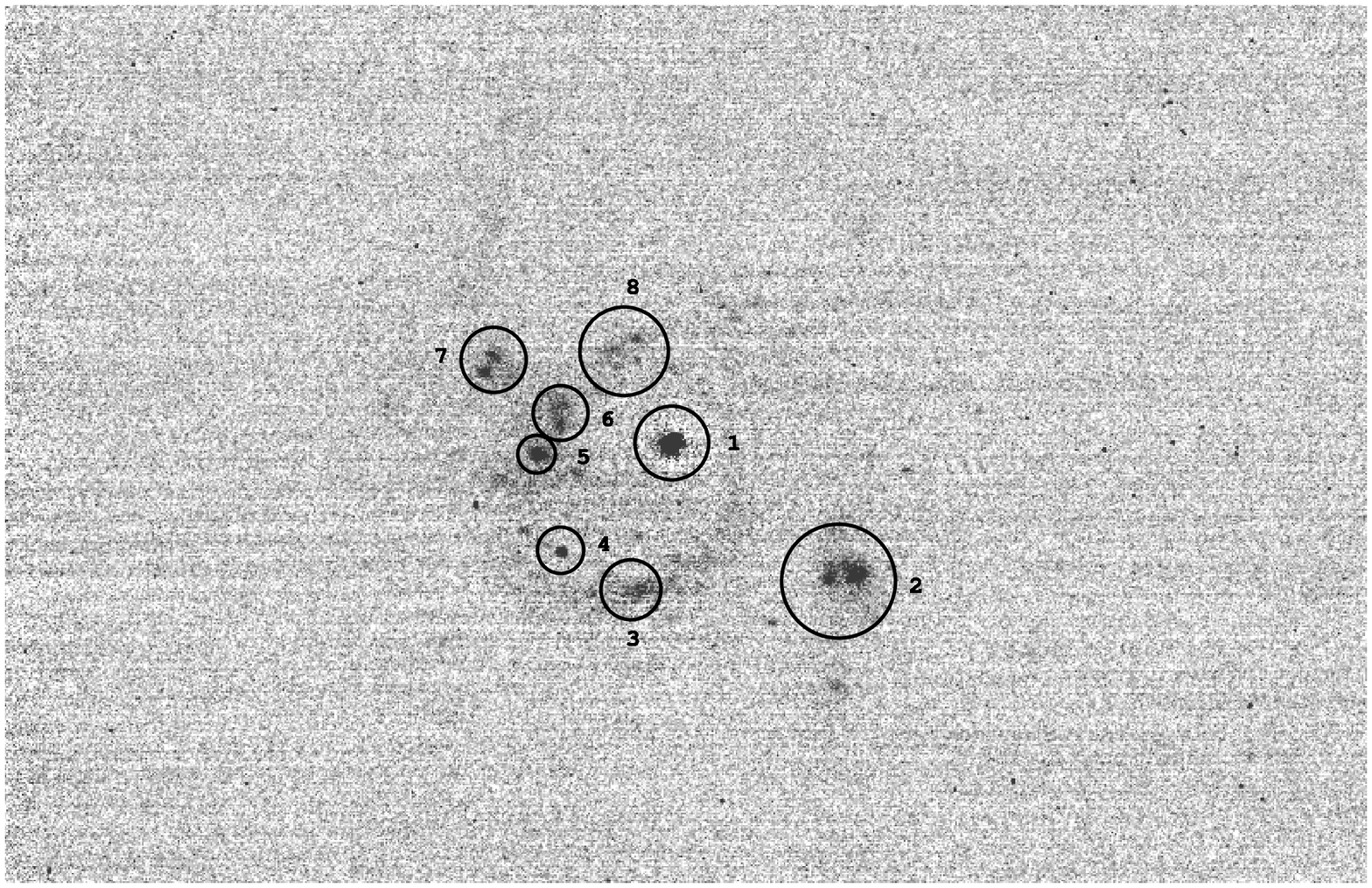}
\includegraphics[angle=0,width=8.5cm]{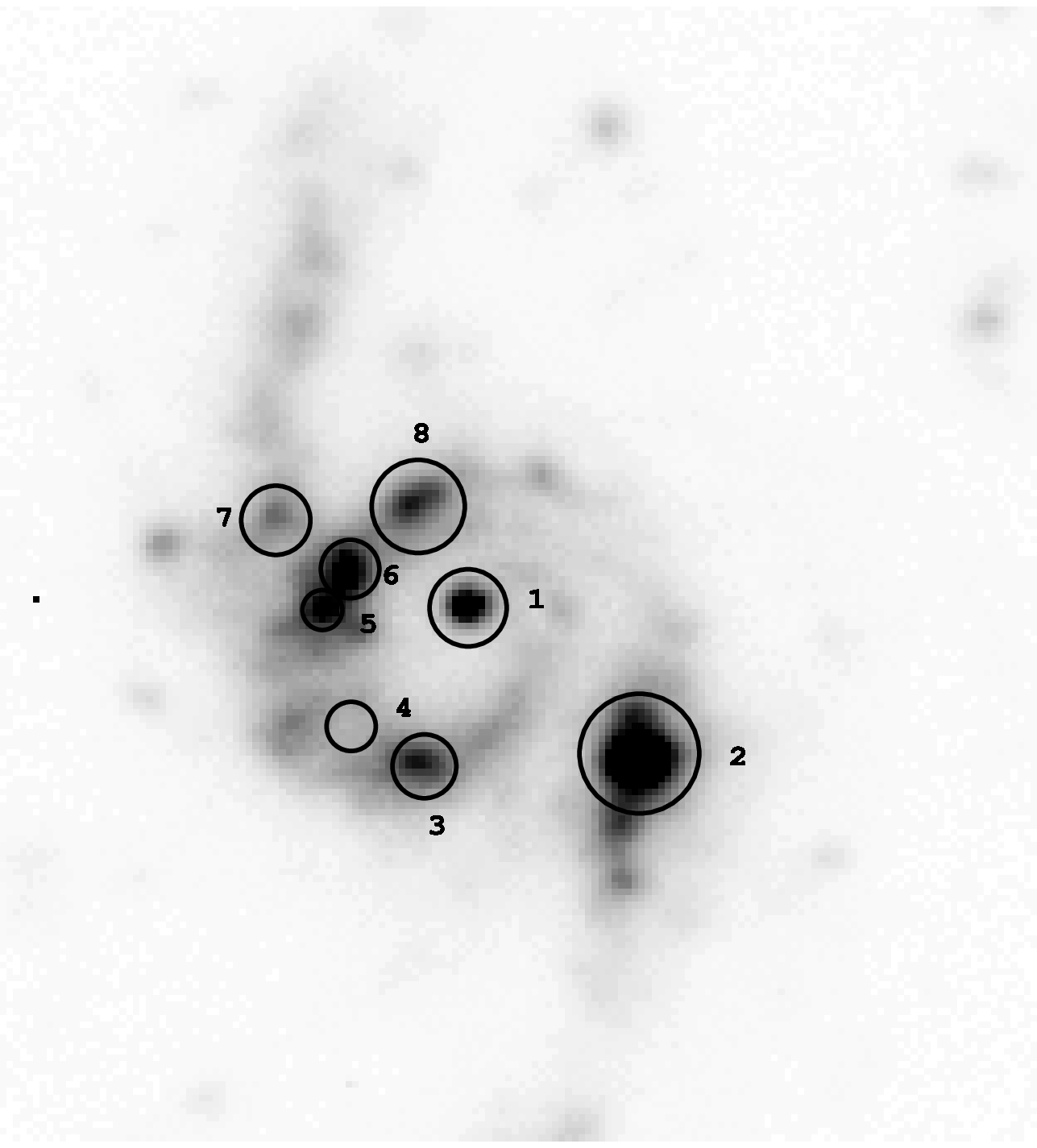}
\caption{Images of UGC~5786 in \Brg\ (left) and \Ha\ (right), showing an area
1\farcm5$\times$1\farcm0, or 7.8$\times$5.0~kpc at the adopted distance
to this galaxy. Regions investigated in Table \ref{tbl:habrflx} are labeled.}
\label{fig:u5786_brg_ha}
\end{figure*}

Out of the 22 galaxies observed, only eight show any sign of \Brg\
detections after continuum subtraction.  In the cases of six of these,
there are 1-3 detections of small, isolated regions.  The \Brg\ image
of UGC~5786, however, contained a detection of nearly the whole galaxy,
with eight strong regions that particularly stand out (Fig.
\ref{fig:u5786_brg_ha}, left image).  UGC~5786 is a very strong \Ha\
emitter (Fig. \ref{fig:u5786_brg_ha}, right image), and has the
highest observed \Ha\ flux in the \Ha\ Galaxy Survey.

In order to correlate the \Brg\ regions with their \Ha\ counterparts,
the \Brg\ data were rebinned to match the pixel scale of \Ha\ images.
To measure the coordinate offset between the two sets of data, the
centroid was found for the galaxy nucleus, or a bright non-saturated
star, in both the JKT $R$-band image and the UKIRT $K^{\prime}$-band
image. There was found to be no significant rotation between the two
sets of images.

\Brg\ fluxes were calculated by placing an aperture around each region.
An annular sky region was taken in cases where there were no other
sources nearby.  Random measurement errors were investigated by
changing the size and position of the sky annulus or offset region.
Systematic errors caused by structure in the background were
investigated using 25 artificially-generated sources with a
two-dimensional Gaussian profile, and sizes comparable to the real
sources detected in the \Brg\ frames. For each detected real source,
the artificial sources were scaled to match the real source and
added to the real frame.  Aperture photometry was then performed on
the embedded fake sources and the standard deviation in values was
used as a measure of the error in the determination of the flux of the
real source. These sky-background errors dominate over other sources
of error, such as the uncertainty in the continuum scaling factors.

The apertures around the \Brg\ regions were transformed to
the \Ha\ coordinate system and the \Ha\ flux for the corresponding
region was then calculated.  In most cases, the regions detected in
\Brg\ matched up with obvious regions in the \Ha\ image.  
The errors on the \Ha\ measurements were found to be much less
than those on the \Brg\ measurements.  The dominant
source of error when calculating \Ha\ fluxes was shown in
Paper I to be the uncertainty in the continuum scaling factors
($\sim$10\% for a typical galaxy observed through the h6570 or h6594
narrow-band filters), and the total uncertainty to be between
10 and 15\%.

Table \ref{tbl:habrflx} shows the \Brg\ fluxes and their corresponding
\Ha\ fluxes for all the detected regions, with the exception of a small 
number of weak detections with no \Ha\ counterpart.  The 1$\sigma$
errors given are those derived from the slight movement of the \Brg\
aperture and changing the sky region (col. 5), the variation in counts
measured from fake sources with the same flux as the object (col. 6), and
the error analysis for \Ha\ fluxes in \citet{thesis} (col. 8).
UGC~3711 was observed on two nights: 17/01/01 (N1) and 19/01/01 (N3).
The \Brg\ fluxes for both galaxies are in agreement within the quoted
errors.

In col. 9, the ratio of the detected fluxes is presented.  The
intrinsic flux of the \Brg\ line is 104 times less than that of
\Ha\ and the effective extinction is 7.1 times
lower (following the extinction law of \citealt{cardelli}); thus the
\Ha\ extinction coefficient, $A$(\Ha), is given by :

%\begin{equation}
$$
A({\rm H\alpha}) = 2.910 \log{\left(\frac{104}{0.75 \times
\frac{F_{\rm H\alpha}}{F_{\rm Br\gamma}}}\right)}
- A_{\rm G}({\rm H\alpha}).
$$
\label{eqn:AHa}
%\end{equation}

$A_{\rm G}({\rm H\alpha})$ is the correction for Galactic extinction, as quoted on
the NASA/IPAC Extragalactic Database (NED).  The factor of 0.75
corrects for \NII\ contamination in the \Ha\ flux (see Section
\ref{sec:nii}).  This is the correction recommended by \citet{ken83},
however, if we substitute the value 0.823, suggested by the mean
\Ha/(\Ha\ + \NII-total) ratio for the 5 entire galaxies in Table
\ref{tbl:niiall}, the values of $A$(\Ha) will be slightly lower.  As an
example, the extinction found for region 3 of UGC~5786 will fall from
1.48 to 1.37 if this change is made.  Since we are investigating
individual \HII\ regions here, however, we continue to use Kennicutt's
value (which was also derived for luminous individual \HII\ regions)
in this analysis. This value is also consistent with the
mean correction found in Section \ref{sec:niifindings} for individual
regions, but this does constitute a significant source of uncertainty
in the following analysis.

\begin{table*}[t]
\begin{center}  
\begin{tiny}
\begin{tabular}{llcccccccc}
\hline
\hline
UGC & type & region & $F_{\rm Br\gamma}$(Wm$^{-2}$) & $1\sigma $ &
$1\sigma $ & $F_{\rm H\alpha}$(Wm$^{-2}$) &$1\sigma $ &
$\frac{F_{\rm H\alpha}}{F_{\rm Br\gamma}}$ & $A({\rm H\alpha})$ \cr
\hline
2455     & IBm          & 1 & 6.662(--18) & 10.5\%  & 8.8\%   & 2.501(--16)  & 15\% & 37.5  & 1.1$\pm0.1$ \cr
	 &              & 2 & 1.543(--18) & 36.0\%  & 35.0\%  & 2.629(--18)  & 15\% & 1.7   & 5.0$\pm0.4$ \cr
         & & {\rm H$\alpha$}1 & 2.994(--18) & 25.3\%  & 17.7\%  & 9.650(--17)  & 15\% & 32.2  & 1.2$\pm0.3$ \cr
3711 (N1)& IBm          & 1 & 4.460(--18) & 10.3\%  & 17.0\%  & 2.304(--16)  & 15\% & 51.7  & 1.0$\pm0.2$ \cr
         &              & 2 & 1.376(--18) & 15.7\%  & 28.0\%  & 5.278(--17)  & 15\% & 38.4  & 1.4$\pm0.3$ \cr 
3711 (N3)&              & 1 & 4.154(--18) & 6.2\%   & 15.2\%  & 2.304(--16)  & 15\% & 55.5  & 0.9$\pm0.2$ \cr
	 &              & 2 & 1.017(--18) & 11.2\%  & 48.0\%  & 5.278(--17)  & 15\% & 51.9  & 1.0$\pm0.5$ \cr
5731     & SAab         & N & 3.878(--17) & 11.4\%  & 3.6\%   & 1.190(--16)  & 15\% & 3.1   & 4.7$\pm0.2$ \cr
5786     & SABbc     & 1(N) & 2.306(--17) & 2.3\%   & 2.9\%   & 3.881(--16)  & 15\% & 16.8  & 2.6$\pm0.1$ \cr
	 & (pec)        & 2 & 2.175(--17) & 4.3\%   & 2.9\%   & 1.587(--15)  & 15\% & 72.9  & 0.8$\pm0.1$ \cr
	 &              & 3 & 7.864(--18) & 4.2\%   & 7.6\%   & 3.228(--16)  & 15\% & 41.1  & 1.5$\pm0.1$ \cr
	 &              & 4 & 3.307(--18) & 8.3\%   & 14.1\%  & 1.066(--16)  & 15\% & 32.2  & 1.8$\pm0.1$ \cr
	 &              & 5 & 6.305(--18) & 2.0\%   & 9.8\%   & 1.841(--16)  & 15\% & 29.2  & 1.9$\pm0.1$ \cr
	 &              & 6 & 8.373(--18) & 4.5\%   & 7.6\%   & 3.909(--16)  & 15\% & 46.7  & 1.3$\pm0.1$ \cr
	 &              & 7 & 9.488(--18) & 3.2\%   & 7.6\%   & 2.396(--16)  & 15\% & 25.3  & 2.1$\pm0.1$ \cr
	 &              & 8 & 8.399(--18) & 16.6\%  & 7.6\%   & 5.144(--16)  & 15\% & 61.2  & 1.0$\pm0.2$ \cr
         & & {\rm H$\alpha$}1 & 2.820(--18) & 24.0\%  & 16.4\%  & 1.080(--16)  & 15\% & 38.3  & 1.6$\pm0.3$ \cr
6123     & SBb          & N & 8.291(--18) & 9.1\%   & 11.1\%  & 4.983(--17)  & 15\% & 6.0   & 3.9$\pm0.1$ \cr
7985 	 & SABd         &   & 1.229(--18) & 10.8\%  & 32.6\%  & 9.410(--17)  & 15\% & 76.6  & 0.7$\pm0.4$ \cr

\hline
\end{tabular}
\end{tiny} 
\caption[]{Results from the investigation of \Brg\ emitting regions.  Col. (1) 
gives the UGC number of the galaxy,  col. (2) the Hubble type,
col. (3) identifies the region investigated, col. (4) gives the
\Brg\ flux measured for that region, and col. (7) gives the
corresponding \Ha\ flux.  The notation is such that the decimal
exponent is given in brackets and a value of 1.0(--18) should be read
as 1.0$\times 10^{-18}$.  Cols. (5),(6) and (8) give 1$\sigma$
errors on these measurements, as described in the text.  Col. (9)
gives the ratio of the fluxes and col. (10) gives the \Ha\
extinction coefficient calculated from these values.}
\label{tbl:habrflx}
\end{center}
\end{table*}

The values of $A$(\Ha) for each region are recorded in col. 10.  In
most cases the extinction values lie between 0.5 and 1.8~mag, in
agreement with the findings of previous studies.  Four regions have
significantly higher values ($>$2.5~mag).  The high extinction regions
in UGC~5731, UGC~5786 and UGC~6123 are all situated in the nucleus of
the galaxy, and are indicated with an N in Table \ref{tbl:habrflx}.
The high extinction region in UGC~2455 coincides with a region that is
bright in the $R$-band image, but appears faint and diffuse when
viewed in \Ha. Finally, we note that similar data taken with 4.2~metre
William Herschel Telescope using the INGRID 1024$\times$1024 near-IR
array camera (\citealt{pack03}) show clear \Brg\ emission from the
nucleus of UGC~4779. This emission implies a highly dust-embedded
source at the centre of this galaxy ($A({\rm H\alpha})=$ 6--7), given
the weak \Ha\ emission found from the corresponding region.

%--------------------------------------------------------------------------------
\subsection{Putting limits on \Brg\ detections}

Due to the low number of \Brg\ detections in our observations, we
determined the minimum flux that a region would need in order to give
a 3$\sigma$ detection in each image, thus enabling us to derive useful
limits from non-detections.  This was done by creating a frame with
randomly-placed artificial sources as discussed in Section
\ref{sec:brg_detect}.  The generated frame was then re-normalised and
added to each of the observed \Brg\ galaxy images.  Photometry was
performed on the artificial sources and the mean value, along with the
1$\sigma$ error was calculated.  The re-normalisation factor was
adjusted until the mean value fell below 3$\sigma$.  The mean count
rate was then converted into a flux in the same way as if a real \Brg\
source was being investigated.  This calculated flux corresponds to
the minimum \Brg\ source flux that we could reliably detect (to a
3$\sigma$ certainty) in each frame.
The results varied from frame to frame, but were all between 1.2 and
5.7 $\times 10^{-18}$ Wm$^{-2}$.  This range is around the same as the
values detected for the fainter \Brg\ regions.  Thus we are detecting
\Brg\ at the very limits of what is possible with the UKIRT
images.

We also investigated whether there were any \Brg\ regions which should
have been detected, given the intrinsic line ratios which we assume
here, but which were not seen.  For each galaxy observed by UKIRT, the \Ha\
image was examined and the fluxes of the regions with the brightest
\Ha\ emission were measured.  If an extinction of $A$(\Ha) = 1.1~mag
is assumed, there should be corresponding \Brg\ fluxes approximately
44 times fainter than these
\Ha\ fluxes, and under the most extreme assumption of no extinction, a
\Brg\ source should be present with 1/104 times the \Ha\ flux.  In all
cases where the predicted \Brg\ flux for a region was greater than or
close to the calculated minimum \Brg\ flux that would give a 3$\sigma$
detection, aperture photometry was carried out on that region in the
\Brg\ image.

Fourteen regions were investigated in 9 different galaxies.  In the
majority of cases any \Brg\ flux found was below the 3$\sigma$
certainty level.  Two new regions were identified however, one in
UGC~2455 and a further one in UGC~5786 (below region 2 in Fig.
\ref{fig:u5786_brg_ha}).  Data for these regions are presented in Table
\ref{tbl:habrflx}, with the regions identified as UGC~2455 \Ha\ 1 and 
UGC5786 \Ha\ 1.

\begin{table*}[t]
\begin{center}  
\begin{tiny}
\begin{tabular}{llcccc}
\hline
\hline
UGC     & region & $F_{H\alpha}$(Wm$^{-2}$) &  $F_{Br\gamma}$(Wm$^{-2}$) & $\frac{F_{H\alpha}}{F_{Br\gamma}}$ & $A(H_{\alpha})$ \cr
\hline
2392    &        & 1.240(--16)         & $<$1.299(--18)        & $>$95.46                   & $<$0.20 \cr
2455    & a      & 6.540(--17)         & $<$1.759(--18)        & $>$37.18                   & $<$1.06 \cr
        & b      & 7.260(--17)         & $<$1.759(--18)        & $>$41.27                   & $<$0.93 \cr
3711    &        & 1.300(--16)         & $<$1.802(--18)        & $>$72.14                   & $<$0.59 \cr
5731    &        & 1.600(--16)         & $<$2.693(--18)        & $>$59.41                   & $<$1.00 \cr      
5786    &        & 1.070(--16)         & $<$1.742(--18)        & $>$61.42                   & $<$0.97 \cr
6797    &        & 1.530(--16)         & $<$2.162(--18)        & $>$75.39                   & $<$0.70 \cr
7232    &        & 1.130(--16)         & $<$2.611(--18)        & $>$43.28                   & $<$1.39 \cr
7690    &        & 1.980(--16)         & $<$3.092(--18)        & $>$64.04                   & $<$0.89 \cr
7985    & a      & 1.380(--16)         & $<$2.727(--18)        & $>$50.61                   & $<$1.20 \cr
        & b      & 1.190(--16)         & $<$2.727(--18)        & $>$43.64                   & $<$1.39 \cr
        & c      & 1.820(--16)         & $<$2.727(--18)        & $>$66.74                   & $<$0.85 \cr

\hline
\end{tabular}
\end{tiny} 
\caption[]{Extinction limits for the regions with the brightest \Ha\ fluxes, but with no 
\Brg\ detections above the 3$\sigma$ limits calculated using randomly 
placed artificial sources.}
\label{tbl:extlimits}
\end{center}
\end{table*}

The extinction values obtained here are consistent with those found
previously.  For the remaining regions, the upper limits for
detectable \Brg\ fluxes found using the randomly-placed artificial
sources were used to place upper limits on the extinction coefficient
for these regions.  These are shown in Table \ref{tbl:extlimits}.  The
limits show that the extinction values in these \Ha\ selected \HII\
regions do not lie above the range of extinction values found by other
authors (i.e. 1.8~mag).  The region investigated in UGC~2392 has
$A(H\alpha) < 0.20$~mag and, thus, appears to contain much less extincting
dust than normal.

From our study of artificial sources, no 3$\sigma$ limits for
\Brg\ detections were found with fluxes less than the lower limit value of 
$F_{\rm H\alpha}/104$.  This value is the minimum limit in the case where
there is no extinction, and this gives confidence in our photometry,
reduction procedures and the physical assumptions made.

%-------------------------------------------------------------------------------

\subsection{\Brg\ fluxes in the literature} 

\Brg\ fluxes for a further two galaxies in our \Ha\ sample have been 
measured previously during spectroscopic studies of extinction in
starburst galaxies.  These values have been published in
\citet{kawara}, \citet{ho}, and \citet{calzetti}.

The \Ha\ fluxes for the corresponding areas were found from our data
and used to calculate the values for the \Ha\ extinction coefficient
in these regions.  In the case of UGC~12699, the region investigated
was the bright starburst nucleus.  For UGC~8098, \citet{calzetti}
observed the main star-forming region, which is located a long way
south of the nucleus.

The quoted \Brg\ fluxes, the measured corresponding \Ha\ fluxes and
the calculated values for the \Ha\ extinction coefficient are
presented in Table \ref{tbl:brg_lit}.

\begin{table*}[t]
\begin{center}  
\begin{tiny}
\begin{tabular}{cccccc}
\hline
\hline
Reference       & UGC   &   $F_{Br\gamma}$(Wm$^{-2}$)  & $F_{H\alpha}$(Wm$^{-2}$)   &  $F_{H\alpha}/F_{Br\gamma}$ & $A(H_{\alpha})$ \cr
\hline
\cite{kawara}   & 12699 &   $5.1\pm1.2$(--17)   & $2.23\pm0.33$(--15) &  43.7                & 1.32$\pm0.26$  \cr
\cite{ho}       & 12699 &    4.7(--17)          & $1.95\pm0.29$(--15) &  41.5	             & 1.38$\pm0.17$  \cr
\cite{calzetti} & 12699 &   $4.99\pm0.20$(--17) & $2.23\pm0.33$(--15) &  44.7                & 1.29$\pm0.05$  \cr
\cite{calzetti} & 8098  &   $3.05\pm0.18$(--17) & $2.72\pm0.41$(--15) &  89.2	             & 0.53$\pm0.07$  \cr

\hline
\end{tabular}
\end{tiny} 
\caption[]{Results using \Brg\ fluxes quoted in the literature.}   
\label{tbl:brg_lit}
\end{center}
\end{table*}

The three values for UGC~12699 are all in very good agreement within
the quoted uncertainties.  The extinction value in UGC~8098 is low, but
within the range quoted in the literature for individual \HII\
regions.
%-------------------------------------------------------------------------------

\subsection{Investigating extinction through \Brg\ observations - 
summary and conclusions}

Twenty-two galaxies were observed through \Brg\ filters at UKIRT in
order to investigate the effects of extinction internal to some of the
galaxies observed in the \Ha\ Galaxy Survey sample.  Only 8 of the
galaxies provided any detections of \Brg\ however, with most of these
containing just 1-3 isolated regions of measurable flux.  UGC~5786 was
the only object where nearly the whole galaxy could be detected in
\Brg.  These data cannot be used, therefore, to investigate the
overall effective extinction for entire galaxies.

\begin{figure}
\centering
\rotatebox{-90}{
\includegraphics[height=8.8cm]{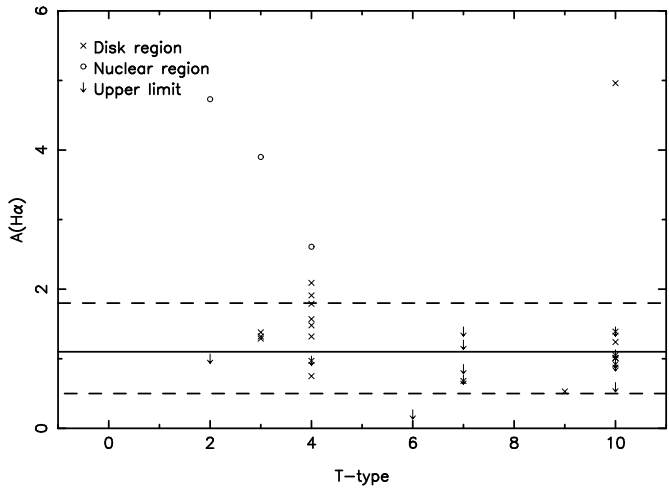}
}
\caption{The \Ha\ extinction coefficients and upper limits calculated
using \Brg\ fluxes, plotted against Hubble T-type.  Error bars have
been omitted for clarity, but uncertainties can be found in Tables 
\ref{tbl:habrflx} and \ref{tbl:brg_lit}.  The 
solid line is positioned at $A$(\Ha)=1.1~mag, and the dashed lines at
0.5 and 1.8~mag.}
\label{fig:AHa_v_T}
\end{figure}

\Ha\ extinction coefficients were calculated for each of these
individual regions and have been plotted in Fig. \ref{fig:AHa_v_T},
along with the upper limits derived from investigating the regions
with the highest \Ha\ fluxes.  In most cases, $A$(\Ha) values are found
to be in the 0.5-1.8~mag range quoted in the literature and indicated by
the dashed lines in the plot.  Error bars have been omitted here for
clarity, but one third of the plotted values are consistent with the
\citet{ken98} value of 1.1~mag, represented by the solid line.  
Higher extinction values were found in nuclear regions, indicating the
presence of large amounts of extincting dust.

Insufficient data were available here to investigate any relationship
with galaxy morphology.  It is clear, however, that $A$(\Ha) varies even
within a single galaxy, and hence a significant fraction of the
extinction must be associated with the star-forming regions
themselves. In the following Section, we will use published data from
the literature to continue our investigation into internal extinction.

%__________________________________________________________________

\section{Internal extinction from literature \Ha\ and \Hb\
observations.}
\label{sec:ucm}

\subsection{The UCM \Ha\ survey}

The UCM survey (\citealt{zam94, zam96}) used a Schmidt telescope, a
low-dispersion objective prism and photographic emulsion to search a
wide field of sky for low redshift emission line galaxies (ELGs).  The
UCM instrumentation limits the survey to z $\lesssim$ 0.045.
\citet{gallego95} report the detection of 264 ELGs in an area covering
471.4 square degrees, more than half of which (138 objects) do
not appear in any published catalogue.

Follow up imaging has been published by \citet{vitores1, vitores2},
papers which contain the Hubble type data used here, and
statistical analyses of the sample galaxies, respectively.  The
morphological distribution of the sample is dominated by late-type
galaxies (66\% being Sb or later) with $\sim$10\% presenting typical
parameters of E-S0 types, and a further $\sim$10\% being irregulars.
This result is a consequence of the UCM selection by the presence of
emission features.  Nine blue compact dwarf galaxies were also
detected.  Follow-up slit spectroscopy was obtained, enabling the
calculation of \Ha\ and \Hb\ fluxes (\citealt{gallego96}).

The UCM dataset thus provides galaxy ellipticities (\citealt{perezg01}),
morphologies (\citealt{vitores1}) and $E(B-V)$ colour excesses
(\citealt{gallego96}) for a large number of galaxies.  These can be used
to investigate the dependence of internal extinction on galaxy
inclination and morphology, which proved impossible with our
Br$\gamma$ observations.

The UCM line fluxes were obtained spectroscopically from slit widths
between 2 and 4~arcsec.  The mean redshift of the galaxies is
approximately 0.02, thus the derived $A$(\Ha) values will be
predominantly for the central regions. There are significant
uncertainties in line fluxes for weak-lined galaxies due to the 0.3~nm
correction applied to compensate for photospheric absorption
underlying both \Ha\ and H$\beta$ lines (\citealt{perezg03}).  As a
result, the analysis presented below was done twice; once for all
galaxies with relevant data, and then including only galaxies with
H$\beta$ EW greater than 1~nm and H$\alpha$ EW greater than 6~nm, for
which the absorption-correction uncertainties will be minimised.

%----------------------------------------------------------------------------
\subsection{The dependence of internal extinction on inclination}

\begin{figure}
\centering
\rotatebox{-90}{
\includegraphics[height=8.8cm]{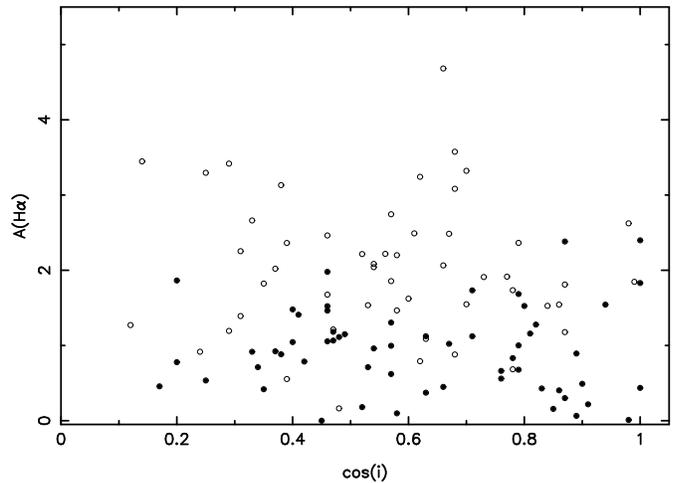}
}
\caption{The \Ha\ extinction coefficient plotted against the cosine of 
the inclination for 109 UCM galaxies.  Solid points show galaxies with H$\beta$
equivalent widths greater than or equal to 1~nm and H$\alpha$
equivalent widths greater than or equal to 6~nm, open points show
galaxies with either or both lines below these limits.}
\label{fig:ext_v_cosi}
\end{figure}

The colour excess values in \citet{gallego96} are computed using
either H$\gamma$/H$\beta$ or \Ha/H$\beta$ observed intensity ratios.
Corrections for Galactic extinction can be found in \citet{vitores1}.
These have been determined from the \citet{burstein82} maps at the
Galactic coordinates of each UCM object.  The corrected values of
$E(B-V)$ can be converted to $A$(\Ha) using the extinction law of
\citet{cardelli}, which gives:
$$ A({\rm H\alpha}) = 0.828 \times R_V \times E(B-V) $$ where the extinction
parameter $R_V$ is defined as $$ R_V \equiv
\frac{A(V)}{A(B)-A(V)} = \frac{A(V)}{E(B-V)} = 3.1. $$ Figure
\ref{fig:ext_v_cosi} shows the calculated \Ha\ extinction internal to
each of the UCM spiral galaxies, in magnitudes, plotted against the
cosine of the inclination for 109 UCM spiral galaxies.  
A least-squares regression method to all 109 points gives $$
A({\rm H\alpha}) = 1.712 - 0.440 \times \cos(i). $$ The linear
correlation coefficient for this fit is 0.106, however, indicating that the
significance for the relationship is just 72\%. A fit to just the
points with strong line emission (the solid points in Fig.
\ref{fig:ext_v_cosi}) gives no correlation at all; the slope is
-0.030, with a correlation coefficient of just 0.012.  It is clear
that selecting on strong emission lines excludes many of the galaxies
with high internal extinction, and the mean A(${\rm H\alpha}$) value is
only two-thirds as large for this subset. The scatter in extinction
values is large compared to the systematic corrections implied by the
above regression fit ($\sim$0.1-0.2 mag). No inclination corrections
were applied to \Ha\ fluxes presented in Paper I.

%----------------------------------------------------------------------------
\subsection{The dependence of internal extinction on morphology}

\label{sec:morphex}

We can also use the UCM data to investigate whether there is
significant type dependence of extinction values.  \citet{vitores1}
classify each of the UCM galaxies according to the classical Hubble
types, as well as including the BCD (Blue Compact Dwarf) type.  They
define BCDs as galaxies possessing all of the following properties:
compact appearance in the direct image, linear size ($D_{24}$) lower
than 10~kpc, luminosity $M_R > -19$, and photometric parameters
typical of later Hubble types.  Spectral information from
\citet{gallego96} was also used for confirmation.

\begin{figure*}
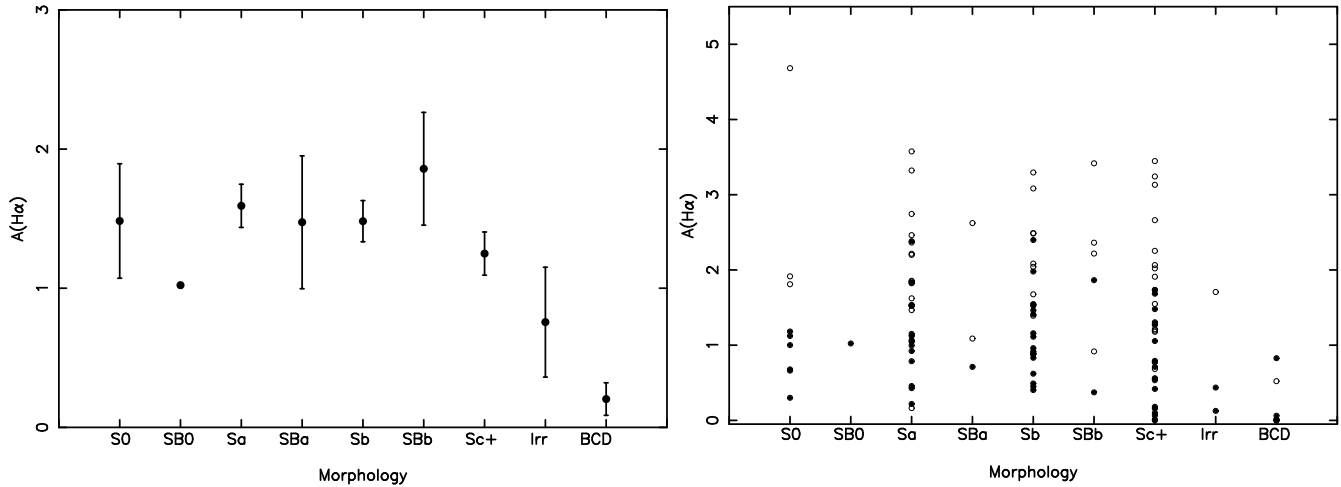

\centering
\rotatebox{-90}{
\includegraphics[height=8.8cm]{A_Hamean_v_type.ps}
}
\rotatebox{-90}{
\includegraphics[height=8.8cm]{fig6b_new.ps}
}
\caption{The \Ha\ extinction coefficient plotted against Hubble type.  
The plot on the left shows the mean values for each type, with the
error bars representing the standard deviation divided by the square
root of the number of galaxies of that type.  The plot on the right
shows each of the 120 UCM galaxies for which both extinction and
morphology data are available. Solid points show galaxies with H$\beta$
equivalent widths greater than or equal to 1~nm and H$\alpha$
equivalent widths greater than or equal to 6~nm, open points show
galaxies with either or both lines below these limits.}
\label{fig:ext_v_type}
\end{figure*}

The extinctions calculated from the colour-excess values in
\citet{gallego96} are plotted against the assigned morphologies in
Fig. \ref{fig:ext_v_type}.  The plot featuring the individual galaxy
values shows a large scatter within each galaxy type.
\citet{vitores1} estimate the typical uncertainty in their adopted
morphologies to be about one Hubble type.  However, this is unlikely to
be the primary cause of the scatter, as inspection of the mean
extinctions for each galaxy type reveals that $A$(\Ha) varies
little between S0 and Sb type galaxies.  The mean extinction is
slightly lower for late-type spirals and drops substantially for
irregulars and BCD galaxies.

\citet{ken98} recommends an \Ha\ extinction value of 1.1~mag for all
galaxy types, while \citet{tresse98} and \citet{tresse02} use \Ha
/\Hb\ ratios to derive a mean value consistent with 1.0 mag in $A_V$,
i.e $\sim$0.75 mag at \Ha .  Figure \ref{fig:ext_v_type} would
suggest higher corrections, of around 1.5~mag, for early-type spiral
galaxies (S0-Sbc), 1.2~mag for late-type spirals (Sc+), and a lower
value of around 0.4~mag for Magellanic irregulars and dwarf galaxies.

This type dependence is generally consistent with results from
previous studies, most of which have been based on broad-band
photometry. \citet{valentijn} found extinction effects to be larger
for Sb and high surface brightness Sc galaxies than for later types.
\citet{han} found systematically larger extinction for types Sbc--Sc
($\gamma_I=$0.90) than for either earlier types, S0/a--Sb
($\gamma_I=$0.73) or later types, Scd$+$($\gamma_I=$0.51).  Finally,
\citet{buat96} found the near-UV 2000\AA\ extinction of Sa-Scd spirals
to be significantly larger than that for Sd-Irr types (0.9 and
0.2~mag. respectively).

It should be remembered that the UCM extinction corrections are
derived from optical lines, and thus there may be a selection effect
against high $A$(\Ha).  However, the values derived are generally
consistent with those from Br$\gamma$/\Ha\ ratios presented in ths
paper, which will be less subject to such effects. The UCM
measurements are also primarily made from nuclear regions and may
somewhat overestimate the extinction characteristic of outer, disk
regions. Figure \ref{fig:AHa_v_T} demonstrates that nuclear
extinction values can be substantially larger than disk values for the
same galaxy, and \citet{domi00} find clear evidence of higher interarm
extinction values towards the central regions of 18 pairs of
overlapping galaxies.  Thus it is possible that the results shown in
Fig. \ref{fig:ext_v_type} would be different if extinctions were plotted
specifically for disk regions, but at present there is no dataset
available to test this possibility. 

The Kennicutt extinction correction was applied when calculating SFRs in
Paper I for ease of comparability with similar studies.  The
effects of applying the type-dependent corrections suggested above
will be investigated in the following Section, and in further papers.

%__________________________________________________________________

\section{Comparison of star formation rates derived from \Ha\ and far-infrared
luminosities}
\label{sec:sfrcomp}

\subsection{Motivation for a test using {\rm \Ha } and IRAS data}

In this Section we will compare SFRs calculated using both \Ha GS \Ha\
fluxes, and FIR fluxes from IRAS (the Infrared Astronomy Satellite).
There are two main reasons for doing this.  Firstly, we can compare
`direct' measures of star formation, as represented by the detected
\Ha\ flux from a galaxy, with indicators derived from dust re-emission,
i.e. the FIR flux.  This will provide a test of the type-dependent \Ha\ extinction
corrections by comparing against an almost extinction-independent
measure of star formation for our galaxies.  Secondly, this will
provide constraints on any very deeply dust-embedded star formation
component which in principle could be missed entirely by optical and
even near-IR measurements, but would certainly result in FIR
luminosity.  Such components are known to exist, for
example in the ultraluminous infrared sources for which Arp~220 is the
local archetype, but it is of great interest to know whether such
components exist, and at what level, in more normal galaxies.

The IRAS all-sky survey resulted in FIR fluxes for over 30,000 galaxies
(\citealt{moshir}), and these can be used as an indirect tracer of 
star formation.  Complications arise from the uncertainties in the
processes responsible for the heating of the IR-emitting dust.
\citet{dev90} and \citet{dev97} argue that high-mass ionising stars
dominate the dust heating.  If this is the case, then the FIR
luminosity of a galaxy should correlate well with the \Ha\ luminosity,
since this is also the result of the UV radiation field from OB stars.
Others (e.g. \citealt{lp87,buat88,sauvage}) have argued that cirrus
emission from dust heated by the general stellar radiation field,
including old, low-mass stars, is also an important factor.  In this
case, we would expect early-type galaxies, with older overall stellar
populations, to produce higher FIR luminosities than predicted by
\Ha-calculated SFRs.

The large size of the \Ha GS sample, combined with the excellent
coverage of galaxy morphologies and surface brightnesses, provides the
basis for investigating the reliability of using FIR
luminosities to trace high-mass star formation.

%----------------------------------------------------------------------
\subsection{Calculation of star formation rates from IRAS flux densities}

The \Ha GS sample was cross-correlated with the IRAS Faint Source
Catalogue v2.0 (FSC) using a search radius of 60\asec\ around the
galaxy positions quoted in NED.  178 galaxies were found in common.

A convenient conversion from flux measurements in the IRAS bands to a
total FIR flux is given by \citet{helou88}; these values were
converted into FIR luminosities using galaxy distances calculated as
described in Paper I, using the Virgocentric inflow model of
\citet{Schechter}, and an asymptotic Hubble constant of 75
km~s$^{-1}$~Mpc$^{-1}$.

The calibration of $L_{\rm FIR}$ to SFR varies within the literature and
depends on assumptions about the star formation timescale and initial
mass function.  \citet{ken98} gives a calibration calculated for
starbursts with ages less than $10^8$ years.  In more quiescent,
normal star-forming galaxies, the empirical relation found by
\citet{buat96} is the most appropriate, and hence is used here:
\begin{equation}
\text{SFR}\ (\Msolar \text{yr}^{-1}) = 3(2-6) \times 10^{-10} L_{\rm FIR}\ (\Lsolar).
\label{eqn:irsfr}
\end{equation}
This correlation is based on IRAS and UV flux measurements of 152 disk
galaxies.  The values given in brackets indicate the 1$\sigma$
interval found for the FIR to UV luminosity ratio.The \Ha\ SFRs listed
in Paper I are determined using the equation derived by \citet{ken94}:
\begin{equation}
\text{SFR}\ (\Msolar \text{yr}^{-1}) =7.94 \times 10^{-35}  L_{\rm H\alpha}\ (\text{W}).
\label{eqn:hasfr}
\end{equation} 
An internal extinction correction of 1.1 mag. $A$(\Ha) was
initially used for all galaxy types in calculating \Ha\ SFR values.

%----------------------------------------------------------------------
\subsection{Comparing \Ha GS and IRAS star formation rates}
\label{sec:ha_v_iras}

\begin{figure}
\centering
\rotatebox{-90}{
\includegraphics[height=8.8cm]{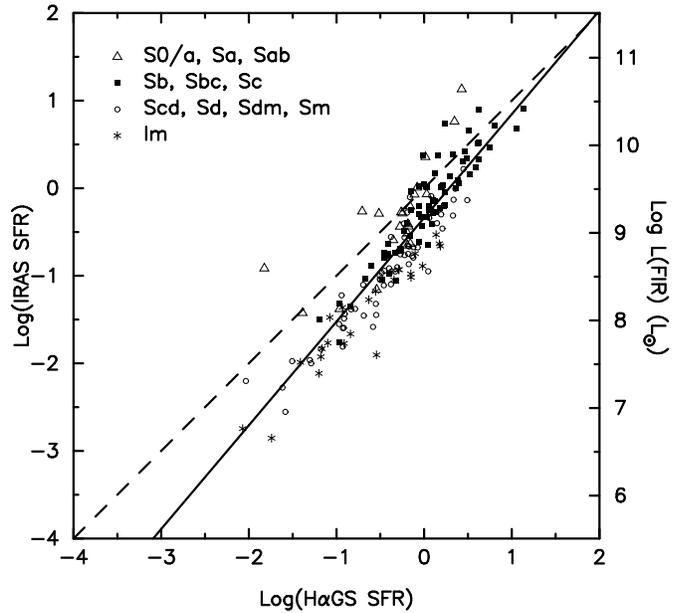}
}
\caption{Comparison of SFRs derived from FIR and \Ha\ data.  
The dashed line is the one-to-one relation and the
solid line is the best-fit line to the data.  The axis on the right
displays the FIR luminosity of the galaxies in Solar units.}
\label{fig:hags_v_iras}
\end{figure}

The 178 galaxies with both \Ha GS and IRAS SFRs are plotted in Fig.
\ref{fig:hags_v_iras}.  The galaxies are coded by galaxy type.  The
dashed line in the figure represents a one-to-one correlation.  The
solid line is the best-fit line calculated from a least-squares
regression fit to the data.  The equation of this line is:
\begin{equation}
\log({\rm IRAS~SFR}) = 1.18\log({\rm H\alpha\ SFR}) - 0.34.
\label{eqn:IR-Ha_line}
\end{equation}
The plot shows a good correlation between the two measures of SFR,
although there is a significant deviation from the one-to-one line.
The linear correlation coefficient is 0.91 (where a value of 1 is a
perfect correlation and a value of 0 indicates no correlation), with a
significance of $>$99.99\%.  For the most rapidly star-forming
galaxies, predominantly Sb-Sc types, the two measures of SFR agree
very well.  For the galaxies with lower SFRs, comprised mainly of the
later types, the FIR tracer seems significantly to underestimate the
star formation, compared to the \Ha.

Figure \ref{fig:hags_v_iras} can be compared with the middle plot of
fig. 1 in \citet{cram}, where \Ha\ SFRs are plotted against SFRs
from radio-power measurements.  The latter can be replaced by
60~$\mu$m IR emission, since the top plot shows a strong one-to-one
correlation between the two.  The trend for IR
measurements to predict lower SFRs at the low end of the scale than
the \Ha\ measurements is visible here.  One possible reason for this
is that galaxies with low SFRs may have lower dust opacities, and thus
be less efficient at absorbing the stellar-UV radiation field from new
stars and re-emitting it at infrared wavelengths.  

The slope of a best fit-line to the plot of \Ha\ SFR vs radio SFR in
\citet{cram} is much less than 1, and lower than the gradient for the
\Ha GS-IRAS relation, 0.85.  Both investigations show that the
assumption of a linear correlation between FIR luminosity and SFR in
Equation \ref{eqn:irsfr} with a slope of unity in the log-log plot
does not appear to be valid.  This is further supported by
\citet{sauvage}, who find a slope of 0.69 when examining the same
relationship for a sample of 135 galaxies.  They argue that this
supports the two-component model where the FIR luminosity comes from
both star-forming regions and quiescent cirrus-like regions of the
ISM.

The right-hand axis in Fig. \ref{fig:hags_v_iras} displays the FIR
luminosity.  The plot clearly shows a strong, tight correlation
between FIR luminosity and the SFR as determined by the \Ha\
luminosity.  If the latter is assumed to be an accurate and reliable
measure of the total star formation (bearing in mind the uncertainties
in extinction and \NII\ corrections), then Equation
\ref{eqn:IR-Ha_line} can be used to derive a non-linear, empirical
correlation between FIR luminosity and the star formation rate of a
galaxy:
\begin{equation}
\text{SFR}\ (\Msolar \text{yr}^{-1}) = 1.6 \times 10^{-8} L_{\rm FIR}^{0.85}\ (\Lsolar).
\label{eqn:new_irsfr}
\end{equation}
Since this equation is derived from SFRs calculated using Equation
\ref{eqn:hasfr}, it implicitly assumes the same Salpeter IMF.

%---------------------------------------------------------------------

\subsection{A test of morphology-dependent extinction corrections}
\label{sec:testcor}

We will now make use of the extinction independence of FIR emission
to perform a simple test of the efficiency of the morphology-dependent
extinction correction derived from the UCM data in Section
\ref{sec:morphex}.

The galaxies in Fig. \ref{fig:hags_v_iras} are coded by
morphological type.  It can clearly be seen that the early-type
galaxies (S0/a-Sab) tend to lie above the best-fit line, indicating an
excess of FIR flux compared to the predictions of the new calibration.
This is in agreement with the findings of \citet{buat96} and supports
the theories of old, low-mass stars contributing significantly to the
dust-heating stellar radiation field.  \citet{young96}, for their
sample of 120 disk galaxies, and \citet{dev90}, for 124 spiral
galaxies from the survey of \citet{kk83}, on the other hand, find no
significant difference in the regions where the early- and late-type
spiral galaxies are located on their equivalent plots.

\begin{figure*}
\centering
\includegraphics[width=8.5cm]{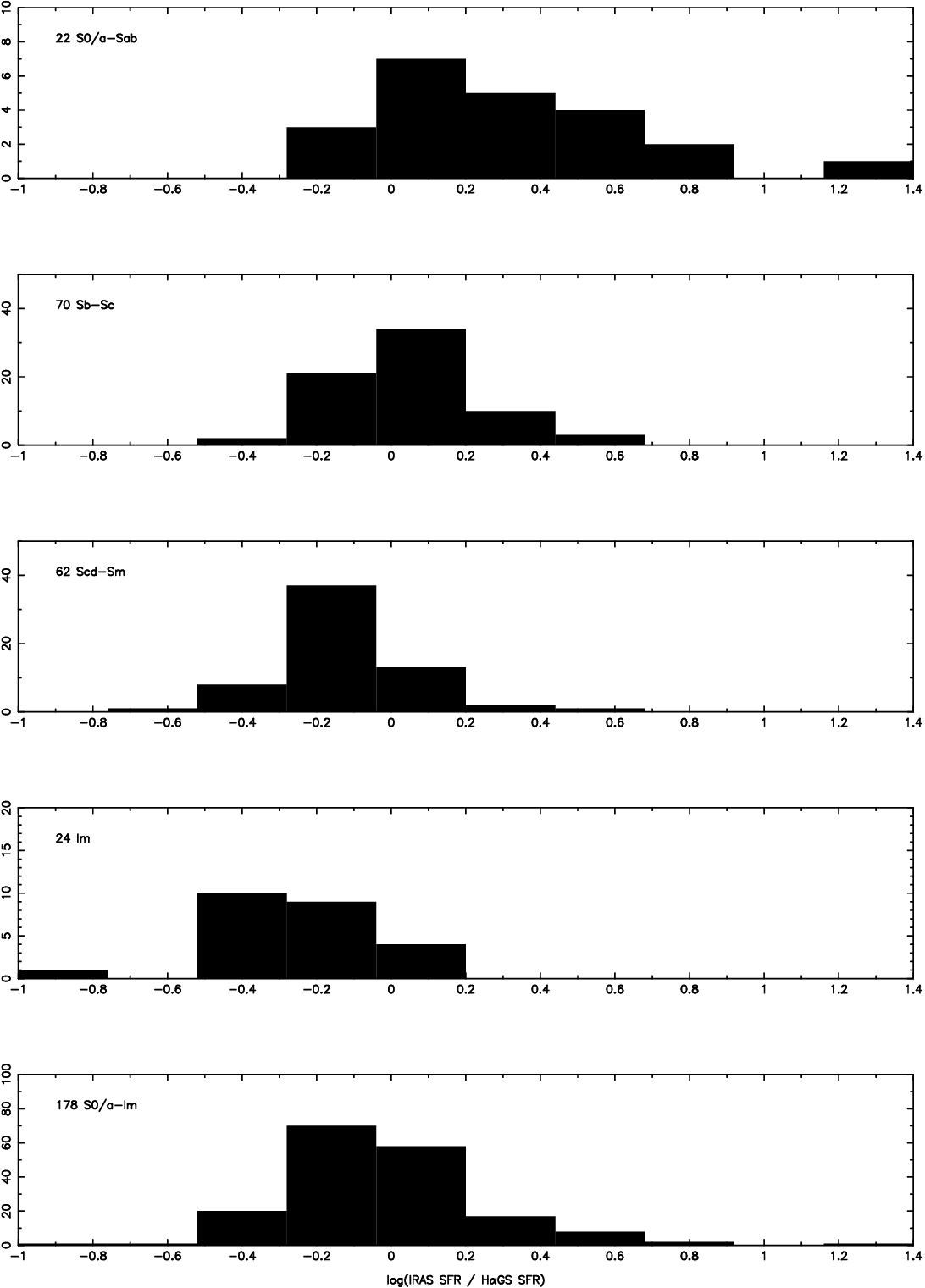}
\includegraphics[width=8.5cm]{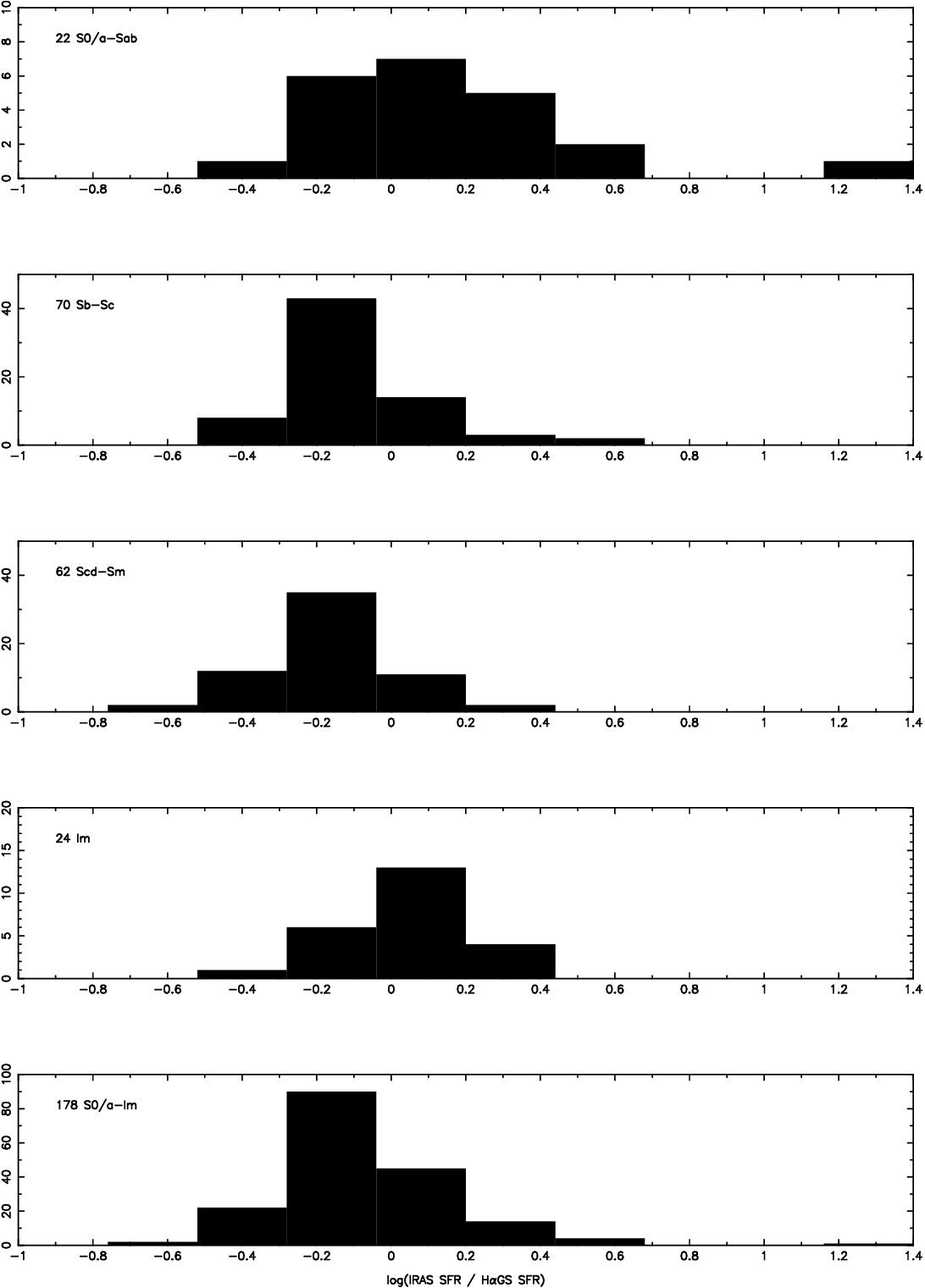}
\caption{Histograms of the ratio of IRAS SFRs to \Ha GS SFRs as a 
function of morphological type. \Ha\ fluxes have been corrected using 
a uniform extinction correction of 1.1~mag. for the histograms on the
left, and using but with the
morphologically-dependent extinction corrections derived from the UCM
data (see text) for those on the right}
\label{fig:iras_hists}
\end{figure*}

Figure \ref{fig:iras_hists} contains histograms of the ratio of the
FIR SFR (as calculated using Equation \ref{eqn:new_irsfr}) to the \Ha\
SFR for different galaxy morphologies.  The difference between early
and late types is immediately clear.  The 22 early-type spirals
(S0/a-Sab) display the largest scatter, but clearly show the highest
IRAS/\Ha GS SFR ratios.  For later
types, this ratio systematically decreases.  The mean ratios
and their standard errors are given in the top row of Table
\ref{tbl:ratio_means}.  The trend is the same as that seen by
\citet{sauvage}, but much stronger than that found by  \citet{dev90}
and \citet{young96}.

\begin{table*}
\begin{center}
\begin{small} 
\begin{tabular}{lccccc}
\hline
\hline
                                       & S0/a-Sab    & Sb-Sc       & Scd-Sm      & Im          & S0/a-Im     \cr
\hline
mean $\frac{\rm SFR(FIR)}{\rm SFR(H\alpha)}$, $A$(\Ha) = 1.1   & 3.17$\pm$0.91 & 1.29$\pm$0.09 & 0.86$\pm$0.05 & 0.62$\pm$0.06 & 1.28$\pm$0.13 \cr
mean $\frac{\rm SFR(FIR)}{\rm SFR(H\alpha)}$, variable $A$(\Ha)& 2.18$\pm$0.66 & 0.89$\pm$0.06 & 0.78$\pm$0.05 & 1.18$\pm$0.11 & 1.05$\pm$0.09 \cr 
$N$                                      & 22            & 70            & 62            & 24            & 178           \cr
\hline
\end{tabular}
\end{small}
\caption[]{The mean ratios of the SFRs calculated from FIR and \Ha\ data 
for different Hubble types.}
\label{tbl:ratio_means}
\end{center}
\end{table*}

In addition to contributions from the old stellar populations, there are
several other possible causes of the observed decrease in
SFR(IRAS)/SFR(\Ha GS) along the Hubble sequence.  Figure
\ref{fig:ext_v_type} shows that the \Ha\ extinction is higher in
early-type spiral galaxies and falls off towards the latest types.
The \Ha\ luminosities used to calculate the SFRs here have all been
corrected by the same extinction factor ($A$(\Ha)=1.1~mag), irrespective
of galaxy type.  If we have underestimated the extinction in
early-type galaxies, then the \Ha\ luminosities will also be
underestimated and the observed SFR(IRAS)/SFR(\Ha GS) will be too
high.  The converse is true for the late types.  \citet{kew02} compare
IR and \Ha\ SFRs for a sample of 93 galaxies before and after
correcting each galaxy individually for extinction.  Before
corrections they find the same trend as above, but after the
corrections are applied they find that the SFR(IR)/SFR(\Ha) ratios for
early and late types are approximately equal.

The results of replacing the standard 1.1~mag correction in the \Ha GS
sample with the corrections suggested by the UCM data in Section
\ref{sec:morphex}, i.e. 1.5~mag for S0/a-Sc, 1.2~mag for Scd-Sm 
and 0.4~mag for Im galaxies, can be seen in Fig.
\ref{fig:iras_hists} and in the second row of Table  \ref{tbl:ratio_means}.
From these results, simple type-dependent extinction corrections seem
to be an improvement on a global correction.

\begin{figure}
\centering
\rotatebox{-90}{
\includegraphics[height=8.8cm]{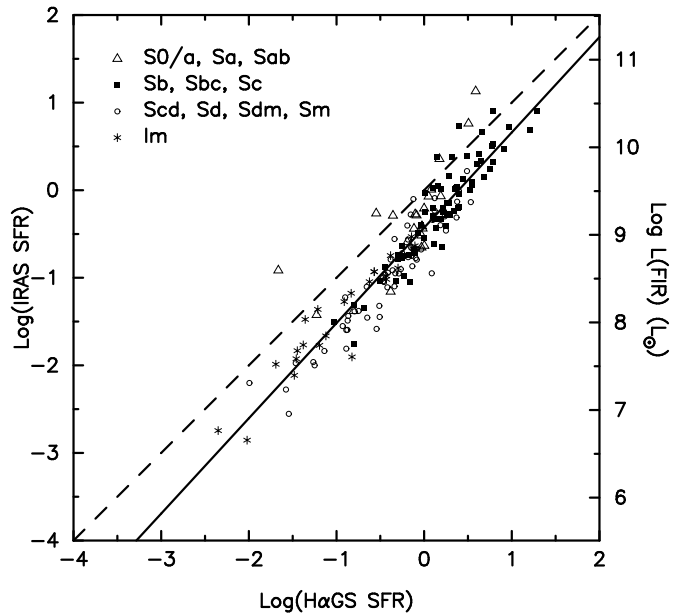}
}
\caption{Comparison of SFRs derived from FIR and \Ha\ data.  
The dashed line is the one-to-one relation and the
solid line is the best-fit line to the data.  The axis on the right
displays the FIR luminosity of the galaxies in Solar units.}
\label{fig:hagscor_v_iras}
\end{figure}

A further point related to the type-dependent extinctions is
that the galaxies with lower optical extinctions may also experience a
lower efficiency of converting stellar light into FIR emission, as
predicted by the models of \citet{char01}, and confirmed
observationally by \citet{hira03}.  This should imply a residual correlation
between SFR(IRAS)/SFR(\Ha ) and galaxy luminosity, even after
correction of \Ha\ fluxes for type-dependent extinction, in the sense
that low-luminosity galaxies will have anomalously low SFR(IRAS)
values.  Also, as mentioned by \citet{hira03}, the SFR(IRAS) values
for luminous early-type galaxies may be boosted by a component due to
dust heating by the strong older stellar population in these galaxies.
Figure \ref{fig:hagscor_v_iras} is a replotting of Fig.
\ref{fig:hags_v_iras}, but with the \Ha\ SFR corrected for
type-dependent extinction, which provides a test of these
predictions. The regression fit to the points has a slope of 1.088,
confirming that the type-dependent extinction corrections have reduced
the difference between the two estimators of SFR, and significantly
reduced the scatter.  However, there is
still a clear underprediction of SFR from FIR fluxes for the lower
luminosity spiral and irregular galaxies; this underprediction becomes less
marked for brighter spiral galaxies, and there is indeed evidence for
excess FIR emission in some of the brighter early-type galaxies.  Thus
Fig. \ref{fig:hagscor_v_iras} gives good qualitative agreement with
the predictions of \citet{char01} and \citet{hira03}.

\subsection{Constraints on deeply embedded star formation in normal
galaxies}

The results described in Section \ref{sec:testcor} also argue against there
being a dominant deeply-embedded star formation component in a
significant fraction of the normal galaxy population sampled by this
study.  Such a population would power a far-IR flux excess,
accompanied by little or no \Ha\ emission, and would result in a tail
of galaxies lying to the right of the main peaks shown in Fig.
\ref{fig:iras_hists}.  Few such galaxies are observed, although some
caveats should be noted.  Firstly, Fig.
\ref{fig:iras_hists} does show a significant fraction of galaxies
with far-IR excesses up to 0.4 dex, or a factor 2.5 relative to the
mean ratio for each type.  This is particularly true for the S0/a--Sab
galaxies. Thus we cannot exclude the possibility that a significant
minority of our galaxies have up to half of their star formation in a
deeply embedded component. Secondly, we do not model the absolute
ratio of far-IR to \Ha\ flux, so it is possible that all galaxies have
a significant embedded star formation component, and for some reason
this is always present in the same ratio to the component that is
visible in the \Ha\ line.  Indeed, \citet{charlot} use modelling of
the \Ha\ and far-IR emission from star forming galaxies to conclude
that \Ha-based measurements of star formation rates underestimate the
true value by a factor $\simeq$3, compared with far-IR measurements.
However, the principal reason they find for this is the absorption of
ionizing photons within star formation regions, rather than the
presence of a dominant embedded component.  It is also interesting to
note that \citet{dopita} find that \Ha\ measurements provide a
reliable estimate of total star formation rates even in luminous and
ultraluminous interacting and merging galaxies, as long as optical
spectroscopic data is available to enable removal of active nuclear
components, and to give corrections for reddening and [N{\sc ii}]
contamination.  Given this, it is not surprising that we have found
good consistency between far-IR and \Ha\ star formation rates for the
more normal galaxies in the present sample, which have had similar
corrections applied.  This lack of dominant deeply-embedded star formation
components has also been demonstrated by the comparison of UV and FIR 
emission by, for example, \citet{buat99} and \citet{hira03}.  It
should be stressed that this has only been demonstrated in the
local Universe and may well not be valid at high redshifts.

\section{Conclusions}
\label{sec:conc}
\renewcommand{\labelitemi}{$\bullet$}

We have studied the two major corrections which need to be applied to
narrow-band \Ha\ fluxes from galaxies in order to convert them to
SFRs, i.e. \NII\ contamination removal and correction for extinction
internal to the galaxy in question.  
\begin{itemize}
\item From an imaging study using
carefully-chosen narrow-band filters, we find that the \NII\ emission
is generally very differently distributed to the \Ha\ emission, and
that in particular nuclear measurements (e.g. from slit spectroscopy)
can significantly over-estimate the contribution of
\NII\ to total narrow-band fluxes.  In most star formation regions in
galaxy disks, the \NII\ fraction is small or negligible, and \NII\
corrections applied in most previous studies may significantly
under-estimate disk star formation rates as a result.
\item We estimate the extinction towards star formation regions in spiral
galaxies from Br$\gamma$/\Ha\ line ratios.  The main results from this
study are that extinctions are larger for regions in galaxy nuclei
compared with those in disks; disk extinction values are similar to
those derived from optical emission-line ratios; and there is no
evidence for heavily dust-embedded regions emerging in the near-IR,
which would be invisible at \Ha.  However, the numbers of galaxies and
individual regions detected using this method are small, and we thus
exploit optical emission-line data for the UCM study to derive global
\Ha\ extinction values as a function of galaxy type and inclination.
\item We find evidence supporting the overall size of corrections
($\sim$1~mag.) applied by previous authors, and determine
typical extinctions to be smaller for late-type dwarfs than for spiral
types.  
\item We show that the application of type-dependent
extinction corrections derived here significantly improves the
agreement between star formation rates calculated using \Ha\ fluxes
and those from far-IR fluxes as measured by the IRAS satellite.  
\item These findings support the idea that heavily dust-embedded
star formation, which would be underestimated using the \Ha\
technique, is not a dominant contributor to the total star formation
rate of most galaxies.
\end{itemize}

%__________________________________________________________________
\begin{acknowledgements}
The Jacobus Kapteyn and William Herschel Telescopes are operated on
the island of La Palma by the Isaac Newton Group in the Spanish
Observatorio del Roque de los Muchachos of the Instituto de
Astrof\'\i sica de Canarias.  The United Kingdon Infrared Telescope is
operated by the Joint Astronomy Centre on behalf of the U.K. Particle
Physics and Astronomy Research Council. This research has made use of
the NASA/IPAC Extragalactic Database (NED) which is operated by the
Jet Propulsion Laboratory, California Institute of Technology, under
contract with the National Aeronautics and Space Administration.
The referee, Dr V. Buat, is sincerely thanked for her many helpful
comments and suggestions.
\end{acknowledgements}
%__________________________________________________________________
%\begin{thebibliography}{}
%\bibliographystyle{bibtex/aa}
%\bibliographystyle{aa}
%\bibliography{refs}

%\end{thebibliography}
\end{document}